\documentclass[sigconf, nonacm]{acmart}
\usepackage{tikz}
\usepackage{subcaption}
\usetikzlibrary{fadings, arrows.meta}

\usepackage{float}
\usepackage{subcaption}
\usepackage{multirow}
\usepackage{enumitem}
\usepackage{booktabs}
\usepackage[linesnumbered,ruled,vlined]{algorithm2e}

\ifx\nocomment\undefined
\newcommand{\laxman}[1]{{\color{brown}{\bf Laxman:} #1}}
\newcommand{\kuba}[1]{{\color{violet}{\bf Kuba:} #1}}
\newcommand{\lars}[1]{{\color{blue}{\bf Lars:} #1}}
\newcommand{\tobias}[1]{{\color{purple}{\bf Tobias:} #1}}
\newcommand{\richard}[1]{{\color{magenta}{\bf Richard:} #1}}
\newcommand{\raj}[1]{{\color{cyan}{\bf Raj:} #1}}
\else
\newcommand{\laxman}[1]{{}}
\newcommand{\kuba}[1]{{}}
\newcommand{\lars}[1]{{}}
\newcommand{\tobias}[1]{{}}
\newcommand{\richard}[1]{{}}
\newcommand{\raj}[1]{{}}
\fi

\newcommand{\hide}[1]{}

\newcommand{\mP}{\mathcal{P}}

\newcommand{\mX}{\mathcal{X}}

\newcommand{\Cmax}{\ensuremath{C_{max}}}
\newcommand{\Cmin}{\ensuremath{C_{min}}}
\newcommand{\Psamp}{\mathcal{P}_{samp}}
\newcommand{\lMax}{\ensuremath{\ell_{max}}}
\newcommand{\AggrFan}{\ensuremath{f_i}}

\providecommand{\norm}[1]{\lVert#1\rVert}
\newcommand{\dist}[2]{{\norm{#1,#2}}}

\newcommand{\emp}[1]{\emph{\textbf{#1}}} 
\newcommand{\myparagraph}[1]{\vspace{0.2em}\noindent\emp{#1.} \,}

\SetCommentSty{mycommfont}

\makeatletter
\patchcmd{\@algocf@start}
  {-1.5em}
  {0pt}
  {}{}
\usepackage{cleveref}
\crefname{section}{Sec.}{Sec.}
\crefname{theorem}{Thm.}{Thm.}
\crefname{lemma}{Lem.}{Lem.}
\crefname{corollary}{Col.}{Col.}
\crefname{table}{Tab.}{Tab.}
\crefname{algorithm}{Alg.}{Alg.}
\crefname{figure}{Fig.}{Fig.}
\crefname{fact}{Fact}{Fact}
\Crefname{table}{Tab.}{Tab.}
\crefname{problem}{Problem}{Problem}

\newcommand{\ourmethod}{PiPNN\xspace}
\newtheorem{defn}{Definition}

\newcommand{\HashPrune}{\textsc{HashPrune}\xspace}
\newcommand{\RobustPrune}{\textsc{RobustPrune}\xspace}

\DeclareMathOperator*{\argmax}{argmax}
\DeclareMathOperator*{\argmin}{argmin}

\newcommand{\mstdeg}{\ensuremath{m}\xspace}

\newcommand{\clustersize}{\ensuremath{C}\xspace}

\setcopyright{none}
\begin{document}

\title[\ourmethod: A Framework for Ultra-Scalable Graph-Based
Nearest Neighbor Indexing]{\ourmethod: Ultra-Scalable Graph-Based \\
Nearest Neighbor Indexing}

\author{Tobias Rubel}
\affiliation{%
  \institution{UMD}
  \city{College Park}
  \state{Maryland}
}
\email{trubel@umd.edu}

\author{Richard Wen}
\affiliation{%
  \institution{UMD}
  \city{College Park}
  \state{Maryland}
}
\email{rwen1@umd.edu}

\author{Laxman Dhulipala}
\affiliation{%
  \institution{UMD and Google Research}
  \city{College Park}
  \state{Maryland}
}
\email{laxman@umd.edu}

\author{Lars Gottesb\"uren}
\affiliation{%
  \institution{Google Research}
  \city{Zürich}
  \state{Switzerland}
}
\email{}

\author{Rajesh Jayaram}
\affiliation{%
  \institution{Google Research}
  \city{New York City}
  \state{New York}
}
\email{}

\author{Jakub {\L}{\k a}cki}
\affiliation{%
  \institution{Google Research}
  \city{New York City}
  \state{New York}
}
\email{}

\begin{abstract}
The fastest indexes for Approximate Nearest Neighbor Search (ANNS) today are also the slowest to build: graph-based methods
like HNSW and Vamana achieve state-of-the-art query performance but have prohibitively large construction times due to relying on
random-access-heavy beam searches. In this paper, we introduce \textbf{PiPNN} (Pick-in-Partitions Nearest Neighbors), an ultra-scalable
graph construction algorithm that avoids this ``search bottleneck'' that existing graph-based methods suffer from.

PiPNN's core innovation is \textsc{HashPrune}, a novel online pruning algorithm which dynamically maintains sparse collections of edges. \textsc{HashPrune} enables PiPNN to partition the dataset into overlapping sub-problems, efficiently perform bulk distance comparisons via dense matrix multiplication kernels, and stream a subset of the edges into \textsc{HashPrune}. \textsc{HashPrune} guarantees bounded memory during index construction which permits PiPNN to build higher quality indices without the use of extra intermediate memory. 

Our extensive experimental study demonstrates that PiPNN builds state-of-the-art indexes up to $11.6\times$ faster than Vamana (DiskANN) and up to $12.9\times$ faster than HNSW. We show that these improvements extend to downstream tasks, yielding speedups of up to $1.9\times$ for approximate $k$-NN graph construction. PiPNN is significantly more scalable than recent algorithms for fast graph construction. PiPNN builds indexes at least $19.1\times$ faster than MIRAGE and $17.3\times$ than FastKCNA while producing indexes that achieve significantly higher query throughput. 
PiPNN  enables us to build, for the first time, high-quality ANN indexes on billion-scale datasets in under 20 minutes using a single multicore machine.
\end{abstract}

\maketitle

\section{Introduction}

High-dimensional vector embeddings are a fundamental datatype used in modern search, information retrieval, classification, and recommendation applications.
For example, vector embeddings are the backbone of a diverse set of applications including entity resolution~\cite{LiLiSuhara:2020,ZeakisPapadakisSkoutas:2023}, retrieval-augmented generation (RAG) systems for generating context for large language models (LLMs) \cite{LewisPerezPiktus:2020, gao2023retrieval}, and recommendation systems \cite{mitra2018introduction, roy2022systematic, XiongXLTLBAO21}.
A fundamental problem in these applications is \textit{nearest neighbor search}: given a query vector (point), find its nearest neighbors from a large set of points
according to some distance function (e.g., $L_2$ or inner product distance).

\begin{figure}
    \centering
    \includegraphics[width=0.95\linewidth]{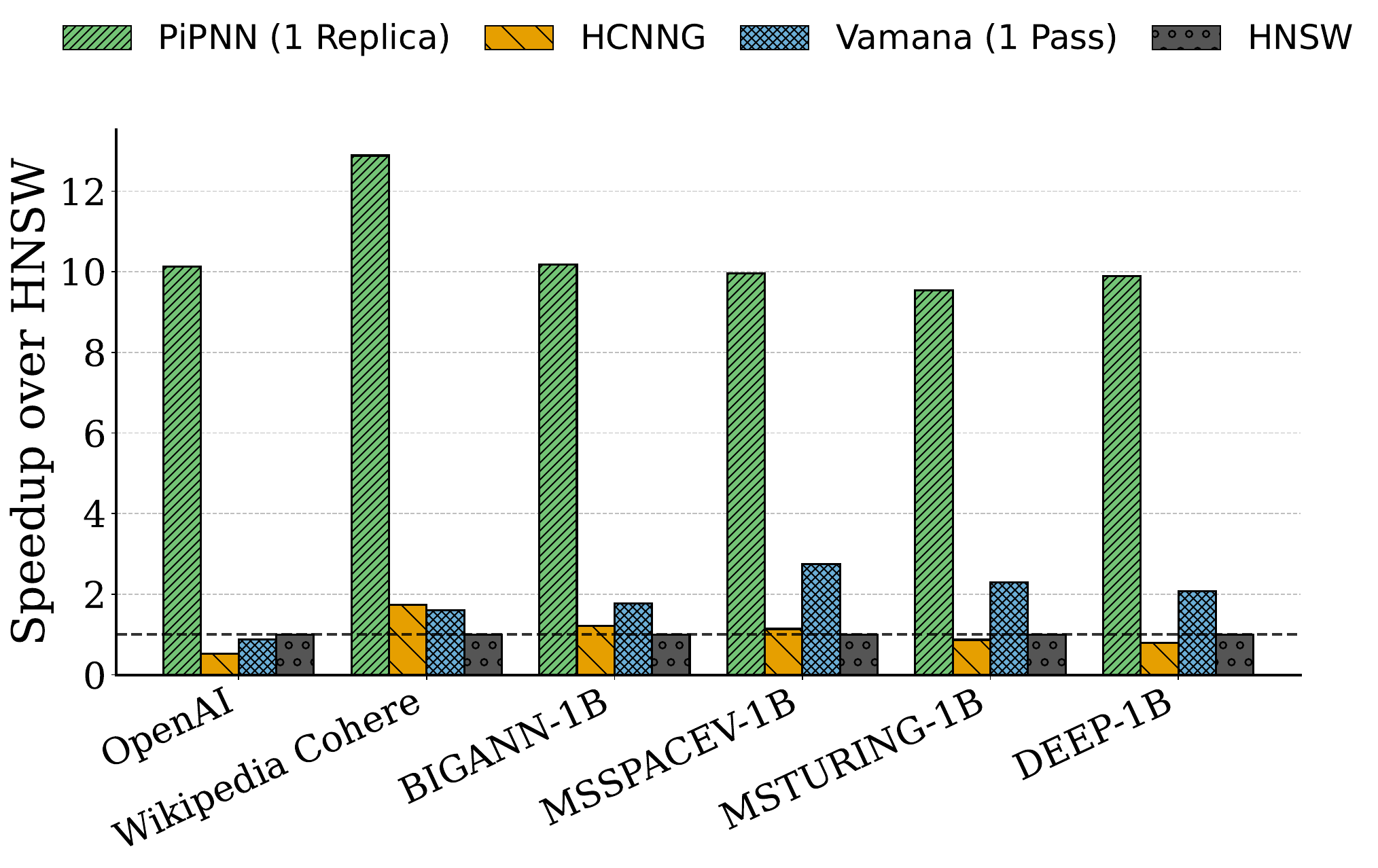}
    \caption{Build time speedup compared to HNSW on six benchmarks, including billion-scale inputs from big-ann-benchmarks.}
    \vspace{-1em}
    \label{fig:teaser}
\end{figure}

Due to the notorious difficulty of finding the exact nearest neighbors in high-dimensional spaces~\cite{beygelzimer2006cover} and the fact that real-world
applications typically tolerate small errors, modern embedding-based applications leverage \emph{approximate nearest-neighbor search (ANNS)}.
In recent years, \emph{graph-based indexing methods} such as HNSW~\cite{malkov2018efficient}, Vamana~\cite{jayaram2019diskann, krishnaswamy2024diskann}, and NSG~\cite{nsg} have become the standard for high-recall and low-latency ANNS.
These methods add edges between nearby points along with carefully chosen long-range edges for navigability, enabling fast queries using \emph{beam search}.

Although graph-based indexing methods are the fastest for querying, \emph{they are also extremely slow to build}, and their
high construction cost undermines their utility in real-world applications where index build time is just as important as query performance.
This occurs, for example, in applications where the index must be rebuilt periodically, or the index parameters must be tuned for new datasets.
Moreover, ANNS often functions as a subroutine within broader algorithmic frameworks, including accelerated $k$-means~\cite{spalding2025scalable}, hierarchical clustering~\cite{yu2025pecann, dhulipala2023terahac}, and approximate $k$-NN graph construction~\cite{carey2022stars}. It is also central to constructing high-dimensional objects like spanners, minimum spanning trees, and single-linkage clustering~\cite{andoni2023sub, beretta2025approximating, azarmehr2025massively, bateni2017affinity}. 
In these contexts, the index construction time counts directly toward the primary metric to optimize, namely end-to-end execution time.
Lastly, the slow construction time of existing graph-based methods presents a severe limitation for large-scale use cases that must index billions of points.

The root cause is that existing graph-based methods rely on incremental insertion, where each point requires a beam search on the partial index to find its neighbors.
This incremental construction approach has been shown to produce very high quality graphs, and has been used in all state-of-the-art graph-based indexes, including HNSW~\cite{malkov2018efficient} and Vamana~\cite{jayaram2019diskann}.
However, the use of beam search during construction leads to a significant volume of random memory accesses, leading to poor pipelining and frequent cache misses.
As a result, incremental graph-based indices suffer from the {\bf \emph{search bottleneck}} where the majority of time spent during index construction is spent performing beam searches, leading to construction times on the order of many hours (and even days) for billion-scale datasets, even on many-core, high-bandwidth machines.
This situation presents the following open problem:

\begin{center}

\emp{Can we obtain state-of-the-art index quality search graphs
while circumventing the search bottleneck inherent to existing
graph-based nearest neighbor search algorithms?}

\end{center}

In this paper we resolve this problem by departing significantly from the approach used by previous state-of-the-art graph indexing methods. 
We instead combine a novel algorithm for maintaining high-quality sparse adjacency lists across batches of candidate edge insertions ({\HashPrune{}}) with {partitioning-based approaches} to efficiently identify candidate neighbors. 
Our graph indexing algorithm is called \emp{\ourmethod{}} (for \emp{Pick-in-Partitions Nearest Neighbors}, pronounced `Pippin'),
a new ultra-scalable algorithm for building ANNS indexes based on 
(1) partitioning the underlying points into a collection of small-sized overlapping partitions called \emph{leaves} via dense ball carving,
(2) building sparse nearest neighbor graphs within each leaf, so as to produce potential candidate edges for the search graph, and 
(3) leveraging an online pruning algorithm (\emp{\HashPrune{}}) to balance both the nearness of edges in the graph as well as directional diversity.

\ourmethod{} circumvents the search bottleneck by eliminating search from the graph-building process altogether.
Instead, we design \ourmethod{} so most of its work is spent in computing all-pairs distances within the leaf partitions using highly optimized general matrix multiply (GEMM) operations using the Eigen library \cite{eigenweb}.
This algorithmic feature of \ourmethod{} enables us to leverage extremely optimized kernels for matrix-matrix multiplication that benefit from vectorized instructions, and are available on essentially all modern hardware platforms.
As we illustrate in Figure~\ref{fig:teaser}, compared to popular (and at this point, highly optimized) methods like Vamana and HNSW that suffer from the search bottleneck, \ourmethod{} is able to construct state-of-the-art indexes up to 11.6x faster than Vamana (average 6.32x) and up to 12.9x faster than HNSW (average 10.4x).

Compared to recent work on improving the scalability of graph-based index construction, including MIRAGE~\cite{voruganti2025mirage}, FastKCNA~\cite{yang2024revisiting}, and LSH-APG~\cite{zhao2023towards}, \ourmethod{} achieves even greater speedups.
We note that these methods were primarily evaluated on 1-million scale datasets.
By contrast, we provide a thorough evaluation of \ourmethod{} at both the 100 million- and billion-scale and show significant speedups over all prior works, which is precisely the regime where scalable index construction is most critical.
%
%

\noindent In summary, the main contributions of this paper are:
\begin{enumerate}[label=(\arabic*),topsep=0pt,itemsep=0pt,parsep=0pt,leftmargin=15pt]
\item A novel online and history-independent pruning algorithm called \HashPrune{} capable of yielding high-quality sparse search graphs when given noisy candidate lists.

\item The \ourmethod{} algorithm that quickly approximates topological neighborhoods via overlapping partitioning, along with fast methods for picking batches of candidate edges from the clusters for consideration by \HashPrune{}.

\item Experimental results comparing \ourmethod{} with existing state-of-the-art graph-based methods with speedups of up to $11.6\times$ for building graph indexes of identical quality and up to $5.8\times$ across four billion-scale datasets.
    
\end{enumerate}

\section{Preliminaries}\label{sec:prelims}

\myparagraph{Approximate Nearest Neighbor Search (ANNS) Background}
Let $\mX \subseteq \mathbb{R}^d$ be a collection of $n$ vectors (points) in $d$ dimensions. Consider some arbitrary \emph{dissimilarity measure} $\norm{\cdot,\cdot}$ over $\mathbb{R}^d$.

\begin{defn}\label{defn:knns}
(k-nearest neighbor search) Given query point $q \in \mathbb{R}^d$, the k nearest neighbor search over $\mathcal{X}$ produces the set $\mathcal{K} \subseteq \mathcal{X}$ such that, $\max_{p\in \mathcal{K}} \norm{q,p} \le \min_{p\in \mX\setminus K} \norm{q,p}$.
\end{defn}

In this work we consider the task of approximating $\mathcal{K}$. 
As is typical, we evaluate the approximations using recall.

\begin{defn}
    (k@k' recall) Let $\mathcal{K}$ be the set of $k$ nearest neighbors of $q$ in $\mathcal{X}$ as defined in Definition \ref{defn:knns}. Then, given $\mathcal{K}^\prime \subseteq \mathcal{X}$ of size $k^\prime$, the k@k' recall of $\mathcal{K}^\prime$ is $\frac{\lvert \mathcal{K} \cap \mathcal{K}^\prime\rvert}{\lvert \mathcal{K} \rvert}$.
\end{defn}

In this paper we follow the literature (see \cite{aumuller2020ann,simhadri2022results,simhadri2024results,manohar2024parlayann}) in focusing on 10@10 recall, and will use the term \emph{recall} to denote this specific notion. For a set $Q$ of queries, we use recall to denote the \emph{mean recall} over $Q$.

\myparagraph{Graph Based Indexes}
Given a set of points $\mX$, the family of graph indexing methods build a \emph{navigation graph} $G(\mX, E)$, where each point $p \in \mX$ is represented by a vertex in the graph.
The goal is to build the graph to ensure that it is searchable via beam search, which we describe next.

\myparagraph{Beam Search (Algorithm \ref{alg:beam_search})}
Given a navigation graph over points $\mX$, the standard query procedure is a greedy search with intermediate storage (called the beam). Pseudocode for this procedure is given in Algorithm~\ref{alg:beam_search}.
Given a source node $s$ in the graph and a query point $q$, a beam search with beam width $L$ works as follows. 
We maintain the two sets: a set $B$ containing the $L$ points seen that are closest to $q$ (often called the ``beam''), and the set $V$ of all points visited. Visiting a point $p \in B \setminus V$ is done by adding all its unseen neighbors to the beam, then adding $p$ to the visited set.
The algorithm begins by initializing the beam with only the source $s$, as well as an empty visited set. Then, we visit the unvisited node in the beam that is closest to $q$, repeating until every node in the beam has been visited. Because the distance of the points in the beam to the target decreases monotonically, this process will eventually converge, at which point we return the final state of the beam.

\begin{algorithm}
    \caption{BeamSearch}
    \label{alg:beam_search}
    \SetKwData{beam}{$B$}
    \SetKwData{beamwidth}{$L$}
    \SetKwData{visited}{$V$}
    \SetKwData{graph}{$G$}
    \SetKwData{source}{$s$}
    \SetKwData{target}{$q$}
    \SetKwData{points}{$\mX$}
    \SetKwData{curr}{$p$}
    \SetKwFunction{FNeighbors}{Neighbors}
    \KwIn{$\graph = (\points, E)$ a graph over set of points $\points$ fitted with dissimilarity measure $\lVert\cdot,\cdot\rVert$, $\source \in \points$ a source point, $\target$ a target point, $\beamwidth \in \mathbb{N}$ the beam width parameter}
    \KwOut{$\beam \subseteq \points$ the $\beamwidth$ closest points to $\target$ encountered during the search}
    $\beam \gets \{\source\}$\;
    $\visited \gets \varnothing$\;
    \While{$\beam \setminus \visited \neq \varnothing$}{
        $\curr \gets \argmin_{u \in \beam \setminus \visited} \lVert u, \target \rVert$\;
        $\beam \gets \beam \cup \FNeighbors{curr}$\;
        $\visited \gets \visited \cup \{curr\}$\;
        \If{$|\beam| > \beamwidth$}{
            $\beam \gets$ closest $\beamwidth$ points to $\target$ in $\beam$\;
        }
    }
    \Return \beam;
\end{algorithm}

\subsection{Offline Pruning \& Incremental Builds}\label{sec:incremental_graph_indexes}

Most state-of-the-art graph-based ANNS methods, including HNSW \cite{malkov2018efficient}, Vamana \cite{jayaram2019diskann}, and NSG \cite{nsg}, can be characterized as employing two-stage process for each point: (1) identifying a subset of candidate neighbors from the universe of points, and (2) applying a pruning algorithm on those candidates to select a sparse set of edges that enforces \emph{directional diversity}.

Pruning algorithms are typically derived from the construction method of the Relative Neighborhood Graph (RNG), which selects edges based on the proximity of candidates to one another \cite{arya1993approximate}. Insofar as pruning requires doing pairwise comparisons between candidate neighbors, these pruning rules are inherently quadratic in the number of candidates. Consequently, one cannot simply prune the entire universe of $n$ points for every vertex, as the resulting $O(n^3)$ construction time would be prohibitive for large datasets.

Many methods circumvent this issue by way of an \emph{incremental construction paradigm}: for each point being inserted, the algorithm performs a search on the current (partial) index to identify a local collection of candidate neighbors. These candidates are then refined using a pruning kernel, such as the \RobustPrune{} algorithm used in Vamana~\cite{jayaram2019diskann} (see Algorithm~\ref{alg:prune}), the RNG-based rules in HNSW and NSG~\cite{malkov2018efficient, nsg}, or the direction-based pruning strategy in SSG~\cite{fu2021high}. While effective at identifying relevant edges, incremental construction suffers from the \emph{search bottleneck}, where the majority of build time is spent on latency-bound random memory accesses that fail to exploit modern hardware parallelism. 

Recent attempts to achieve faster index construction try to reduce the cost of these searches. MIRAGE~\cite{voruganti2025mirage} aims to improve the quality of the base graph in order to make the searches more efficient. LSH-APG~\cite{zhao2023towards} uses LSH to avoid some distance computations in the search, as well as to pick a closer starting point. FastKCNA~\cite{yang2024revisiting} introduces a 'refinement-before-search' framework. Rather than searching a partial graph to find candidates for pruning, it first applies a relaxed RNG constraint ($\alpha$-pruning) to an approximate KNN graph, yielding faster searches for identifying candidate neighbors. However, these methods do not eliminate searches entirely, and thus still suffer from the associated inefficiencies.

An alternative to the search-based approach is to use some kind of overlapping partitioning, as seen in methods like HCNNG~\cite{munoz2019hierarchical}. 
\emph{Partitioning approaches} identify potential neighbors without searching the graph which allows for more cache-friendly algorithms via batched distance computations. 
However, the curse of dimensionality forces clustering methods to generate a significantly larger universe of candidate points per vertex than search-based methods. 
Storing all these candidates for a final ``offline'' prune would require a massive memory footprint, and the resulting candidate lists would be prohibitively large to prune using existing $O(n^2)$ pruning methods. Moreover, existing partitioning approaches do not consistently produce state of the art ANNS indexes. This is because, absent a viable pruning strategy, HCNNG  combines edges produced in each cluster via union, which yields dense adjacency lists with directional redundancy. 

The ideal solution for a partitioning algorithm would be to prune candidates as they are discovered in each partition. 
However, existing pruning kernels do not lend themselves to online, batched execution. Firstly, each batch of candidates would require a quadratic number of distance computations over the union of the batch and existing partial adjacency list, which is computationally prohibitive. Moreover, if a candidate list is divided into arbitrary batches (e.g., partition by partition), traditional pruning rules lose global coherence. For example, a point $z$ might be pruned in an early batch because of an "intermediary" point $y$. If $y$ is itself later pruned during the processing of another partition, the algorithm no longer has a justification for the pruning of $z$, potentially leading to a fragmented graph structure. Thus, existing pruning methods lack the history-independence required to ensure that the final graph structure remains consistent regardless of the order that candidates are processed. These challenges motivate our development of \HashPrune, which maintains a deterministic and sparse graph across arbitrary collections of candidates while simultaneously guaranteeing a bounded memory footprint and fast updates.

\begin{algorithm}
    \caption{RobustPrune}
    \label{alg:prune}
    \SetKwData{cands}{$\mathcal{N}$}
    \SetKwData{pstar}{$p^*$}
    \SetKwData{E}{$E$}
    \SetKwFunction{FAdd}{Add}
    \SetKwFunction{FRemove}{Remove}
    \KwIn{\cands a collection of candidate points fitted with dissimilarity measure $\lVert\cdot,\cdot\rVert$, points $x \in \mathcal{X}$, $\alpha \in \mathbb{R}$, $R \in \mathbb{N}$}
    \KwOut{$\E \subseteq \mathcal{X}\times\mathcal{X}$}
    \While{$\cands \neq \emptyset$ and $|\E| < R$}{
        $y \gets \argmin_{y \in \mathcal{X}} \lVert x, y \rVert$\;
        \E.\FAdd{$(x, y)$}\;
        \cands.\FRemove{$y$}\;
        \For{$z \in \cands$}{
            \If{$\alpha \cdot \lVert y, z \rVert < \lVert x, z \rVert$}{
                \cands.\FRemove{$z$}
            }
        }
    }
    \Return \E\;
\end{algorithm}

\section{Online Pruning via \HashPrune{}}\label{sec:hashprune}

In this section, we devise a history-independent online pruning
technique which we call {\bf \emph{\HashPrune}} which is integral to
the partitioning approach used in \ourmethod{}.

\myparagraph{Motivation}
As mentioned in Section \ref{sec:incremental_graph_indexes}, pruning
algorithms ensure that each vertex connects to neighbors in many
different directions so that greedy search can quickly
route from any one point to any other.
\HashPrune{} accomplishes this objective via locality-sensitive hashing
(LSH)~\cite{charikar2002similarity} to prune candidate edges which are
likely to be directionally similar to other candidates.
The use of LSH also allows for the use of low dimensional
\emph{sketches} during pruning, which eliminates the need to fetch
high dimensional vectors.
We give pseudocode for the \HashPrune{} algorithm in
Algorithm~\ref{alg:hash}, and formally describe the components of
\HashPrune{} next.

\myparagraph{Residualized Hashing}
\HashPrune{} is initialized by generating a set of $m$ random
hyperplanes $\{\mathcal{H}_1, \ldots, \mathcal{H}_m\}$
through the origin.
For each point $p$, we generate an individualized hash
function $h_p$, which takes a candidate point $c$ and hashes the \emph{residual} of $p$ and $c$:
\begin{equation}
    h_p(c) = \bigoplus_{i=1}^{m} \begin{cases}
      1 & \text{if }   \mathcal{H}_i\cdot (c-p) \ge 0,\\
        0 & \text{if } \mathcal{H}_i\cdot (c-p) < 0. 
    \end{cases}
\end{equation}

Here $\oplus$ designates the concatenation operation.
That is, we produce the $i$-th bit in $h_p$ by checking
which side of the $i$-th hyperplane the residual falls on.

\myparagraph{Insertion Procedure}
To use \HashPrune{} to prune candidate neighbors for a given point $p$, we initialize a
\emph{reservoir} $M$ of size \lMax{} (Line~\ref{hash:reservoir}).
To insert a candidate neighbor $c$, we compute the individualized
hash function $h_p(c)$. If another candidate with the same hash
already exists in the reservoir $M$ (Line~\ref{hash:collidestart}),
then we keep the candidate that is \emph{closer} to $p$
(Line~\ref{hash:comp}).
Otherwise, we insert the incoming candidate into the reservoir, and if
the reservoir is full (Line~\ref{hash:full}), we evict the existing
candidate that is furthest from $p$ (Line~\ref{hash:evict}).
Our next theorem shows that \HashPrune{} can be easily used in settings where candidate neighbors of points are obtained in arbitrary orders due to its history-independent guarantees.

\begin{theorem}
    \HashPrune{} is \emph{history-independent}: Given a hash function $h_p$ for some point $p$, a collection of candidates $C$, and a reservoir size $\ell$, the final adjacency list produced by \HashPrune{} is unique and independent of the insertion order of candidates in $C$. 
\end{theorem}

See Supplement \ref{lem:hashprune-history-independence} for a detailed proof of this result.

\myparagraph{Why \HashPrune{} Works}
For each $p$,
and candidates $c, c^\prime$, the classic theorem for LSH gives us
that $P[h_p(c) = h_p(c^\prime)] = (1 - \frac{\theta}{\pi})^m$, where
$\theta$ is the angle between $c-p$ and $c^\prime - p$
\cite{charikar2002similarity}.
Thus, \HashPrune{} draws probabilistic cones centered on each $p$, and
retains the nearest point in each cone.
Figure
\ref{fig:HashPrune_overview} illustrates how the density of the final
graph is impacted by the number of hyperplanes used for the hashes.

\newcommand{\drawhashgraph}[1]{
    \useasboundingbox (-1.2,-1.2) rectangle (1.4,1.2);
    
    \coordinate (P) at (0,0);
    \fill (P) circle (1.2pt);
    \node[below left, scale=0.9] at (P) {$\mathbf{p}$};

    \def\neighbors{50/0.45/c_1, 160/0.85/c_2, 340/0.65/c_3}

    \foreach \ang/\dist/\name in \neighbors {
        \begin{scope}[rotate=\ang]
            \foreach \step in {0,1,...,50} {
                \pgfmathsetmacro{\currWidth}{#1 * (1 - \step/50)}
                \fill[cyan, opacity=0.015] (0,0) -- (\currWidth:1.0) arc (\currWidth:-\currWidth:1.0) -- cycle;
            }
        \end{scope}
    }

    \foreach \ang/\dist/\name in \neighbors {
        \draw[-{Stealth[length=3pt]}, thin, opacity=0.4] (P) -- (\ang:\dist);
        \filldraw[fill=white, draw=black, thin] (\ang:\dist) circle (1.2pt);
        
        \pgfmathsetmacro{\oppAnchor}{\ang+180}
        \node[anchor=\oppAnchor, inner sep=3pt, font=\small] at (\ang:\dist) {$\name$};
    }
}

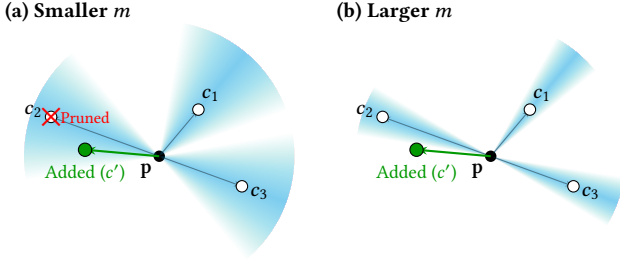
\begin{figure}[t]
    \centering
    \begin{minipage}[t]{0.48\columnwidth}
        \centering
        \begin{tikzpicture}[scale=1.8]
            \drawhashgraph{30} 
            
            \coordinate (C) at (175:0.55);
            \draw[-{Stealth[length=4pt]}, green!60!black, thick] (P) -- (C);
            \filldraw[fill=green!60!black, draw=black] (C) circle (1.4pt);
            \node[green!60!black, below, scale=0.8, yshift=-4pt] at (C) {Added ($c'$)};

            \coordinate (C2) at (160:0.85);
            \node[red, font=\bfseries, scale=1.5] at (C2) {$\times$};
            \node[red, right, scale=0.7, xshift=2pt] at (C2) {Pruned};

            \node[anchor=north west, font=\small\bfseries] at (-1.2, 1.2)
              {(a) Smaller $m$};
        \end{tikzpicture}
    \end{minipage}
    \hfill
    \begin{minipage}[t]{0.48\columnwidth}
        \centering
        \begin{tikzpicture}[scale=1.8]
            \drawhashgraph{11} 
            
            \coordinate (C) at (175:0.55);
            \draw[-{Stealth[length=4pt]}, green!60!black, thick] (P) -- (C);
            \filldraw[fill=green!60!black, draw=black] (C) circle (1.4pt);
            \node[green!60!black, below, scale=0.8, yshift=-4pt] at (C) {Added ($c'$)};

            \node[anchor=north west, font=\small\bfseries] at (-1.2, 1.2)
              {(b) Larger $m$};
        \end{tikzpicture}
    \end{minipage}

    \caption{Impact of resolution on neighbor retention. Shading represents the probability of collision based on number of bits used in the \HashPrune{} hash ($m$). (a) At coarse resolution, $c'$ collides with and evicts the farther neighbor $c_2$. (b) At finer resolution, both edges are retained.}
    \label{fig:HashPrune_overview}
\end{figure}

As opposed to prior pruning algorithms which require $O(\ell)$ comparisons to insert a new point into a pruned adjacency list of size $\ell$, \HashPrune{} can be implemented to both prune and insert a new candidate in $O(\log \ell)$ comparisons so long as the reservoir is not full. When the reservoir is full (i.e. $\ell = \lMax$), \HashPrune{} requires $O(\lMax)$ time to insert a candidate when there is no collision, but otherwise retains the $O(\log \ell)$ behavior for insertion. 
The linear insertion time for non-colliding candidates in a full reservoir results from having to update the furthest neighbor in the reservoir.

\begin{algorithm}
    \caption{HashPrune}
    \label{alg:hash}
    \SetKwData{V}{$\mathcal{V}$}
    \SetKwData{pstar}{$p^*$}
    \SetKwData{E}{$E$}
    \SetKwData{dict}{$M$}
    \SetKwFunction{FGet}{Get}
    \SetKwFunction{FSet}{Set}
    \SetKwFunction{FHasKey}{HasKey}
    \SetKwFunction{FRemove}{Remove}
    \SetKwFunction{FValues}{Values}
    \KwIn{\V a stream of candidate points, fitted with dissimilarity
    measure $\lVert\cdot,\cdot\rVert$, point $p$, size limit $\lMax \in \mathbb{N}$}
    \KwOut{A set $C \subseteq \V$ of size $\min(|\V|, \lMax)$}
    $\dict \gets$ an empty key-value store mapping hashes to points\; \label{hash:reservoir}
    \For{$c \in \V$}{
      \If{\dict.\FHasKey{$h_p(c)$}}{ \label{hash:collidestart}
            \If{$\lVert p, c \rVert < \lVert p, {}$\dict.\FGet{$h_p(c)$}$\rVert$}{\label{hash:comp}
                \dict.\FSet{$h_p(c), c$}\;
              }\label{hash:collidend}
        }
        \Else{
          \If{$|\dict| < \lMax$}{\label{hash:nonfull}
                \dict.\FSet{$h_p(c), c$}\; \label{hash:nonfullset}
            }
            \Else{
                $z = \argmax_{y \in \dict} \lVert p, y \rVert$\; \label{hash:full}
                \If{$\lVert p, c \rVert < z$}{
                  \dict.\FRemove{$z$}\; \label{hash:evict}
                    \dict.\FSet{$h_p(c), c$}\;
                }
            }
        }
    }
    \Return \dict.\FValues{}\;
\end{algorithm}

\myparagraph{Implementation}
For each point $p$, we implement its corresponding key-value store $M$ as an array of capacity \lMax, in which each candidate $c$ in $M$ is represented by its 4-byte point ID, the 2-byte hash $h_p(c)$, and the norm $\lVert p, c \rVert$ stored as a 2-byte floating point number (bf16). Thus, each slot in the reservoir uses a mere 8 bytes. 
This compactness is important to maintaining a reasonable memory footprint, as we keep one reservoir per point.

Instead of using the actual points $p,c$ to compute the residual $c-p$, we pre-compute $m$-dimensional sketches for each point in the dataset by letting $Sketch(v) = [v\cdot \mathcal{H}_i : 0 \le i < m]$. Then $h_p(c)$ can be computed by concatenating the sign of the difference between each dimension of $Sketch(c)$ and $Sketch(p)$. This yields the same hashes but requires loading $m$-dimensional vectors which are one or two orders of magnitude smaller than the $d$-dimensional vectors \ourmethod{} is designed to index.

Candidates in the reservoir are stored in sorted order by hash, letting us quickly binary search for a given hash. When the reservoir is full, we identify the farthest candidate $z = \argmax_{y \in M} \lVert p, y \rVert$ in the reservoir through a simple linear scan, then subsequently cache it. If we find that a new candidate does not warrant an eviction (because its norm is greater than that of $z$), then we can reuse the cached $z$ for the next candidate without a linear scan.

Although one could likely implement a more efficient data structure for the key-value store (e.g. a compact hash table), profiling indicates that maintaining the reservoir typically constitutes no more than 5\% of our total construction time. Thus, unless significant improvements are made to the speed of the rest of the \ourmethod{} algorithm, improving the reservoir scheme is unlikely to yield a notable impact on overall running time.

\section{\ourmethod{} Algorithm}\label{sec:main_algorithm}

\myparagraph{Motivation}
\ourmethod{} uses \HashPrune{} in conjunction with highly optimized partitioning-based approaches to index construction to achieve state of the art index quality in line with that of the best incremental methods. 
\ourmethod{} takes better advantage of modern hardware by 1) partitioning the points into cache-friendly units of work and 2) once vector embeddings are grouped into cache-friendly chunks, compute all pairwise distances within the chunk using highly optimized matrix multiplication kernels.

\myparagraph{Overview of the \ourmethod{} Algorithm}
High-level pseudo-code for \ourmethod{} is given in Algorithm \ref{alg:pipNN}. At a high level \ourmethod{} works as follows: (1) creating an overlapping {\bf \emph{partitioning}} $\mathcal{B} = \{b_1,\dots,b_t\}$ of the points in $\mathcal{X}$ (Line~\ref{line:partition}), (2) generating candidate edges between each point and some subset of the points it resides with via a simple sparsification routine (Line~\ref{line:leafprune}), and (3) using \HashPrune{} to prune those candidate edges for each point (Line~\ref{line:addedges}). 
We are able to improve query performance further by performing a final \RobustPrune{} of the overall candidate list for each $x \in \mathcal{X}$ (Line~\ref{line:finalprune}).


\begin{algorithm}
    \SetKwData{X}{$\mathcal{X}$}
    \SetKwData{BP}{$BP$}
    \SetKwData{G}{$G$}
    \SetKwData{B}{$\mathcal{B}$}
    \SetKwData{E}{$new\_edges$}
    \SetKwFunction{FBucket}{Partition}
    \SetKwFunction{FPrune}{Prune}
    \SetKwFunction{FPick}{Pick}
    \SetKwFunction{FaddG}{Add\_Edges}
    \SetKwFunction{FPaddG}{Prune\_And\_Add\_Edges}
    \SetKwFunction{FPruneNode}{PruneNode}
    \caption{\ourmethod{}}
    \label{alg:pipNN}
    \KwIn{\X a collection of points with dissimilarity measure, \BP a collection of hyper-parameters}
    \KwOut{\G$=(V,E)$ a graph over $\mathcal{X}$}
    G $\gets (\mathcal{X}, \varnothing)$\;
    $\mathcal{B}$ $\gets$ \FBucket{\X,\ \BP}\;\label{line:partition}
    \ForPar{$b_i \in  \mathcal{B}$}{
        $\mathsf{new\_edges}$ $\gets$ \FPick{$b_i$,\ BP}\;\label{line:leafprune}
        G.\FPaddG{$\mathsf{new\_edges}$}\;\label{line:addedges}
    }
    \If{\BP.final\_prune}{
        \ForPar{$x \in$ \X}{
            \FPruneNode{x}\;\label{line:finalprune}
        }
    }
    \Return \G\;
\end{algorithm}

\subsection{Partitioning via Randomized Ball Carving}\label{sec:partitioning}

As previously mentioned, incremental methods seek to find for each point a collection of nearby points by way of greedy searches on the partial index. \ourmethod{} instead seeks to find an overlapping partitioning of the dataset such that nearby points are likely to fall into the same partition. Possible solutions to this problem include locality-sensitive hashing, k-means partitioning, and repeated binary partitioning \cite{indyk1998approximate,bartal2001approximating,munoz2019hierarchical}.
We include an ablation study in Supplement \ref{subsec:partitioning_ablation} which evaluates these choices, but ultimately found that \emph{Randomized Ball Carving} (RBC) achieved the best balance of construction time and index quality.

In a subproblem $\mP \subseteq \mX$, Randomized Ball Carving (RBC) selects $\ell$ randomly chosen points $p_1, \dots, p_\ell \in \mP$ to serve as \emph{leaders}.
It then assigns each point in $\mP$ to its nearest leader, forming $\ell$ subproblems. In \ourmethod{}, we let $\ell$ be the smaller of some percentage of the $\lvert \mP \rvert$ points in a subproblem $\Psamp\cdot \lvert \mP \rvert$ and a hard-cap (typically 1000). 
It then recursively performs ball carving on subproblems with more than \Cmax{} points, until eventually producing leaves $\mathcal{B}_1, \dots, \mathcal{B}_b$ each of size at most \Cmax. Note that a single execution of RBC yields a disjoint partition. This procedure could be run multiple times to produce a collection of partitions which jointly produce an overlapping partitioning, in a process we refer to as \textbf{replication}.
In the following, we propose a more computationally efficient approach. 

\myparagraph{Fanout}
Given that the recursion tree in RBC has arity roughly $\Psamp\cdot n$, we can avoid replicating the partitioning process and instead assign each $x$ to the subproblems associated with its $k$ nearest leaders at the top level. The arity of the tree is large at high levels. 
Thus, a single recursive process can be used to produce an overlapping clustering which is comparable with respect to its utility in index construction (see Supplemental Figure \ref{fig:qps_replication_bar}). 
This fanout optimization yields performance improvements because it replaces the process of replicating the entire recursive process repeatedly (each of which requires a pass over the data for each replicate) with a single recursive process and, therefore, a single pass over the entire dataset at the top level (Supplemental \Cref{sec:fanout_ablation}).

While this modification improves index construction times, we found that delaying some fanout to lower recursion levels additionally aids performance, since the subproblems are smaller (yielding better cache-behavior), the top-k aggregation is cheaper, and the parallel group-by operation to form clusters runs on fewer entries.
We observed that fanout along a roughly geometric sequence (e.g., 10 on the top level, 3 on the second level) produces neighbor lists of similar quality with extremely fast partitioning times.
We refer to this scheme as {\bf \emph{multi-level fanout}}.
This scheme can be used to replace replicating the entire partitioning procedure with no change in index performance (Supplemental Figure \ref{fig:qps_replication_bar}), but improves partitioning time by between $1.35\times$ and $2.5\times$ (as seen in Figure \ref{fig:build_replication_bar}).

To ensure that the partitions are all within bounded sizes which will likely fit into cache during the prune step we define the user tunable maximum cluster size \Cmax  to be fairly small (typically between 1024--2048), and return a partition once it is beneath this size. 
We also define a minimum cluster size \Cmin and merge partitions randomly (never exceeding a cluster size of \Cmax) if they fall beneath this cutoff. 
This ensures that each partition gives enough potential neighbors for the prune algorithm.
Algorithm \ref{alg:Partition} gives pseudocode for this partitioning strategy. 

\begin{algorithm}
   \caption{Randomized Ball Carving}
    \SetKwComment{Comment}{$\triangleright$\ }{}
   \SetKwFunction{FSample}{SampleLeaders}
   \SetKwFunction{Fmin}{min}
   \SetKwFunction{FMerge}{Merge}
   \SetKwFunction{FPartition}{Partition}
   \SetCommentSty{textnormal}
   \Comment{Constants: \Cmax (max cluster size), \Cmin (min cluster size), $\Psamp$ (leader fraction), $\mathsf{fanout}: \mathbb{Z} \to \mathbb{Z}$  (function of depth
   )}
   \label{alg:Partition}
   \KwIn{$\mP \subseteq \mX$ a collection of points with dissimilarity measure, depth$=0$}
   \KwOut{non-disjoint partition $\mathcal{B} = \{b_1,\dots,b_t\}$}
   \If{$\lvert \mP \rvert \le \Cmax $}{\Return{\{$\mP$\}}\;}
       $\mathcal{L} \gets$ \FSample{$\mP$, $\Psamp \cdot |\mP|$}\;
   $\mathsf{local\_fanout}$ $\gets$ \Fmin{$\mathsf{fanout}$(depth),$\lvert \mP \rvert$}\;   
  $\mathcal{B} \gets \{\emptyset,\dots,\emptyset \}$ of size $\mathcal{L}$ \;
   \ForPar{each point $x \in \mP$}{
       \textit{nearest\_leaders} $\gets$  the \textit{local\_fanout} leaders in $\mathcal{L}$ closest to $x$\;
       \For{each leader $l \in \textit{nearest\_leaders}$}{
           Add $x$ to the set in $\mathcal{B}$ corresponding to $l$\;
       }
   }
   \FMerge{$\mathcal{B},\Cmin,\Cmax$}\;
   $\mathcal{B}$' $\gets \emptyset$\;
   \ForPar{$b \in \mathcal{B} : \lvert b \rvert \ge \Cmax$}{
       $\mathcal{B}.remove(b)$\;
      $\mathcal{B} \gets \mathcal{B}\ \cup$ \FPartition{$b$,depth+1}\; 
   }
   \Return{$\mathcal{B}$}\;
\end{algorithm}

\subsection{Leaf Building}\label{sec:pruning}

RBC produces an overlapping partitioning, $\mathcal{B} = \{b_1,\dots,b_t\}$ where each $b_i$ is a collection of up to \Cmax{} points we term ``leaves''. We can obtain distances within a given leaf more efficiently by precomputing a distance matrix small enough to fit in cache, something that would be infeasible on the full dataset and without the fact that we restrict each point's interactions to those other points which share a cluster with it. This allows us to use efficient matrix multiplication kernels~\cite{eigenweb} to compute all-pairs-distances and thus amortize the cost of distance computations within a cluster. 

While at this point it is possible to simply use \HashPrune{} to prune the entirety of the leaf for each point, performance can be improved by first carefully picking a sparser set of candidate edges to prune. Considering too many candidates can result in indexes of excessive density which perform poorly in practice, in addition to adding needless computation which slows down indexing time. This is not a defect unique to \HashPrune{}: In our testing we found both \RobustPrune{} and the RNG prune condition produced degraded indexes when provided with massive candidate lists.

Instead, we pick a subset of the possible edges to add as candidates to the \HashPrune{} structure for each point by building a k-NN graph over the leaf and then bi-directing the edges. We considered other methods as well, including minimum spanning trees, directed k-NNs, inverted k-NNs, and running \RobustPrune{} for each point within the leaf. More detailed descriptions of these other methods are available in Supplement \ref{sec:picking_ablation}, along with an ablation of their relative performances.

\subsection{Final Pruning}\label{sec:pruning}

\HashPrune{} can be used alone to produce high quality indexes by simply taking the graph generated by building k-NN graphs on leaves produced by randomized ball carving. \ourmethod{} also provides the option to post-process the graph with a final \RobustPrune{} on each point. Because \RobustPrune{} has a real-valued $\alpha$ parameter for tuning sparsity, it is more finely tunable to the topology of a given dataset than \HashPrune{} which uses discrete hash buckets to estimate angular cones. This allows for additional sparsification, which we find improves query throughput by a small amount. The size of the adjacency lists provided to \RobustPrune{} by \ourmethod{} are very small relative to those in Vamana\footnote{Our experiments with Vamana showed that due to adding back edges, \RobustPrune{} can be called on candidate lists exceeding 10k elements.}, and thus the cost of the final pruning step is low.

\section{Experimental Evaluation}\label{sec:exps}
\myparagraph{Machines}
Our experiments were carried out on one of two identical machines with four Intel Xeon Platinum 8160 CPUs each (totaling 192 CPUs) and 1.5TB of DRAM.
Our experiments on billion-scale datasets and $k$-NN graph building experiments were run on a Google Cloud c4-highmem-288 machine with 288 vCPUs and 2.2TB of DRAM.
Results in a single plot or chart are always generated on the same machine so as to obviate any differences. 

\myparagraph{Baseline Algorithms \& Parameters} 
We compare our approach against five graph-based algorithms. We evaluated \ourmethod{} against HCNNG~\cite{munoz2019hierarchical}, Vamana~\cite{jayaram2019diskann}, and HNSW~\cite{malkov2018efficient}. Both Vamana and HNSW are widely used due to their excellent query performance. 
We utilize the implementations of HCNNG and Vamana provided by the ParlayANN library~\cite{manohar2024parlayann}. 
To the best of our knowledge, the ParlayANN implementation of Vamana is the fastest and highest quality implementation of a shared-memory graph-based ANNS algorithm available today.
In addition, we evaluate \ourmethod{} against MIRAGE~\cite{voruganti2025mirage} and FastKCNA~\cite{yang2024revisiting}, which aim to achieve high-quality graphs in less build time than Vamana and HNSW. We do not evaluate against LSH-APG~\cite{zhao2023towards}, which was shown in ~\cite{yang2024revisiting} to build indexes slower than FastKCNA.
We also evaluate against SymphonyQG \cite{gou2025symphonyqg}; however, because it is primarily optimized for query speed (and using quantization, which all others methods we test avoid) as opposed to build speed, we include this comparison in \Cref{sec:additional_experiments}.
Regarding configuration, we adopt the optimal hyperparameters identified in ParlayANN for Vamana, while for HNSW,  HCNNG, MIRAGE, and FastKCNA we follow the hyperparameters recommended by their respective authors. To ensure a fair comparison, we restrict the maximum graph degree to 64 for all methods with the exception of HCNNG. For HCNNG, for which the maximum degree is dependent on other parameters, we allow a maximum degree of 90 as recommended by the authors. We were unable to run MIRAGE and FastKCNA on billion-scale datasets because they exceeded the 2.2TB of DRAM available on our machine, and additionally FastKCNA could not be run on WikiCohere due to a lack of support for the MIPS dissimilarity measure. 

\myparagraph{Datasets} We evaluate on four billion-scale datasets and two smaller high-dimensional datasets, accessed through the BigANN Benchmarks framework~\cite{simhadri2024results, simhadri2022results}. Table~\ref{tab:datasets} summarizes the datasets. For ablation studies, we use 100-million sized subsets of BigANN and MS-SPACEV, as well as Wikipedia-Cohere and OpenAI-ArXiv.

\begin{table}[thb]
\centering
\caption{Summary of datasets used in our experiments.}
\label{tab:datasets}
\small
\begin{tabular}{lrrllc}
\toprule
\textbf{Dataset} & \textbf{Size} & \textbf{Dim.} & \textbf{Type} & \textbf{Metric} & \textbf{Cite} \\
\midrule
BigANN & 1B & 128 & uint8 & L2 & \cite{jegou2011product} \\
DEEP & 1B & 96 & float32 & L2 & \cite{babenko2016efficient} \\
MS-SPACEV & 1B & 100 & int8 & L2 & \cite{simhadri2022results} \\
MS-Turing & 1B & 100 & float32 & L2 & \cite{simhadri2022results} \\
Wikipedia-Cohere & 35M & 768 & float32 & MIPS & \cite{reimers2022cohere} \\
OpenAI-ArXiv & 2M & 1536 & float32 & L2 & \cite{neelakantan2022text} \\
\bottomrule
\end{tabular}
\end{table}

\subsection{Optimizations in \ourmethod{}}\label{sec:pipnn_optimizations}

We first describe and evaluate the key algorithmic and implementation optimizations in \ourmethod{} which jointly contribute to the superior index construction times shown in Section \ref{subsec:comparison}. 

\myparagraph{Multi-Level Fanout}
In Section \ref{sec:partitioning}, we explained that \ourmethod{} uses a single recursive process to select multiple leaders per level, referred to as fanout, instead of repeatedly replicating the entire partitioning. We claimed that this significantly speeds up partitioning times. In particular, replication requires us to execute the topmost levels of recursion many times across different runs. On the other hand, fanout allows us to dispense many copies of each point into lower subproblems from only a single execution at the root level. Multi-level fanout lets us partially extend this benefit past the root level to the second level or deeper. See Supplement \ref{sec:fanout_vs_replication} for a more detailed explanation. This yields massive speedups overall, because the large number of leaders per subproblem causes the recursion to terminate after very low depth (often, a mere 2 to 3 steps of recursion suffices). That is, the topmost levels of recursion constitute the majority of the partitioning time.

\begin{figure}
    \centering
    \vspace{-1em}
    \includegraphics[width=\linewidth]{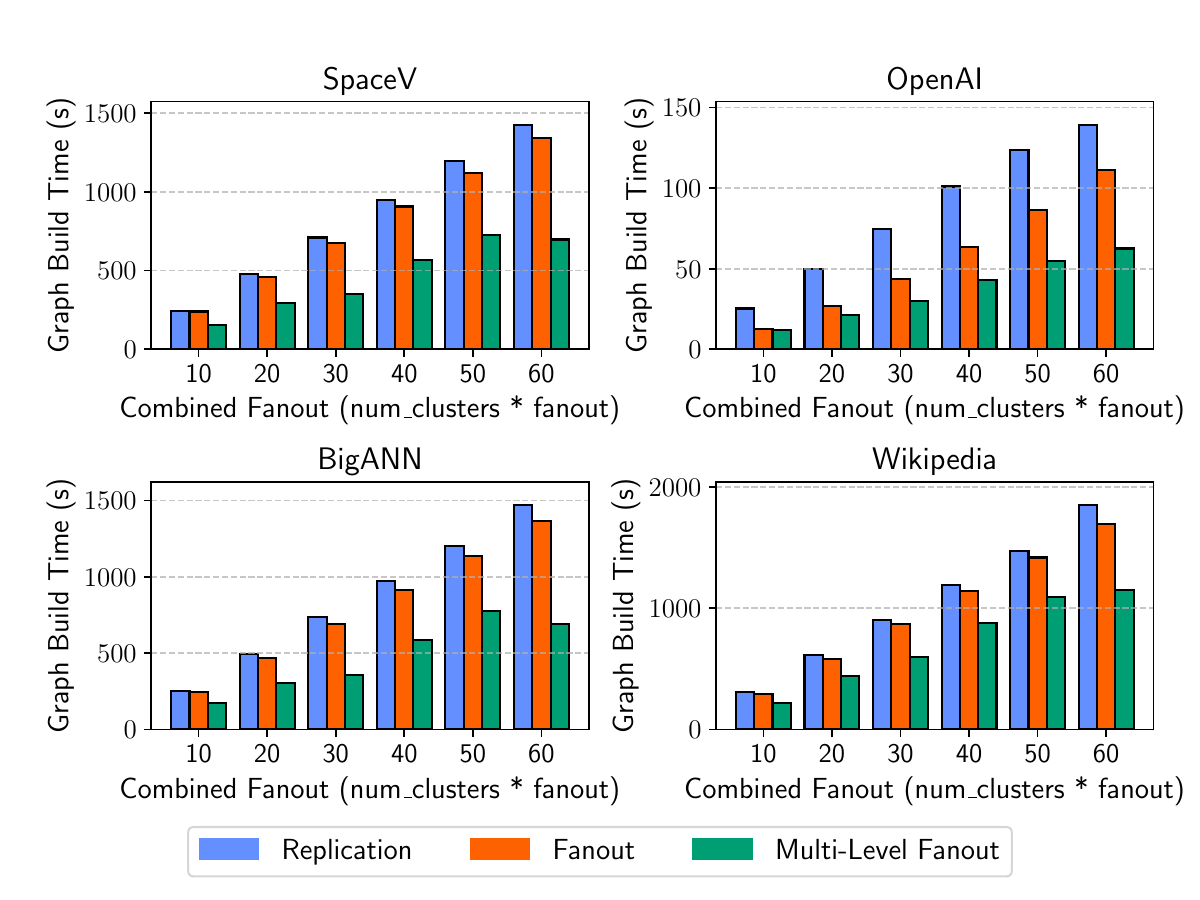}
    \caption{\small Multi-level fanout significantly reduces build times, with the effect becoming more pronounced as fanout is increased.}
    \vspace{-1em}
    \label{fig:build_replication_bar}
\end{figure}

We ablate the construction time yielded by these three strategies. Using fanout instead of replication yields an average $1.47\times$ speedup in partitioning time, and multi-level fanout yields an average $3.35\times$ speedup (see Supplemental Figure \ref{fig:bucket_replication_bar}). Figure \ref{fig:build_replication_bar} shows how these speedups impact total construction time. Further, we find that replication, fanout, and multi-level fanout are equivalent with respect to the quality of the resultant graph (Supplemental Figure \ref{fig:qps_replication_bar}).

\myparagraph{Leaf-Building Optimizations}
A key benefit of the partition-based approach of \ourmethod{} is that it enables the use of optimized matrix multiplication and sorting routines to quickly prune candidate lists. 
In particular, we compute a distance matrix containing all distances between points in a leaf via the Eigen library~\cite{eigenweb}.
We further optimize leaf building by obtaining the nearest neighbors of each point in a leaf using a vectorized partial sort implementation from the Highway library \cite{google_highway}. 
In total, these optimizations combine to yield a $27\times$ speedup in leaf construction time. 
For a more detailed treatment of these optimizations, see Supplement \ref{sec:leaf_optimization_ablation}.

\myparagraph{Parameter Tuning \HashPrune{}}
We evaluated the effect of using more or fewer hash functions in HashPrune. Using fewer hash functions makes collisions more likely, thus in effect broadening the angle of difference that two points must have relative to the source point. On the other hand using more hash functions could quickly make collisions very unlikely even for nearby points. We evaluated the performance of \ourmethod{} indexes built with $b \in \{6,8,10,12,14,16\}$ hash functions. We saw that a broad range of values are acceptable, with only the use of 6 hash functions degrading performance, as displayed in Supplemental Figure \ref{fig:hashbitspacev}. We chose to use 12 as the default for our experiments.

\begin{figure}
    \centering
    \includegraphics[width=\linewidth]{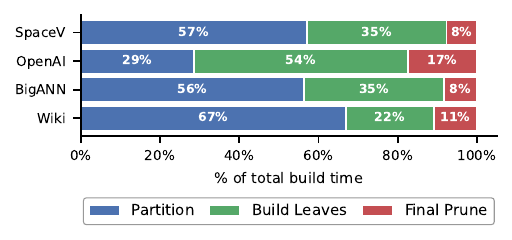}
    \caption{Portion of time spent in Partitioning, Leaf Building, and Final Prune for the four ablation datasets.}
    \label{fig:time_breakdown}
\end{figure}

\myparagraph{Running Time Breakdown}
For reference, we provide a breakdown of the time spent in various phases of \ourmethod{}, namely partitioning into leaves (Partition), building $k$-NN graphs over those leaves and passing the edges into \HashPrune{} (Build Leaves), and performing a prune on the final \HashPrune{} reservoirs (Final Prune). We show the percentage of time taken by each phase in Figure \ref{fig:time_breakdown}. 

\begin{figure*}[t]
	\begin{subfigure}[b]{.3\textwidth}
		\centering
		\includegraphics[width=\textwidth]{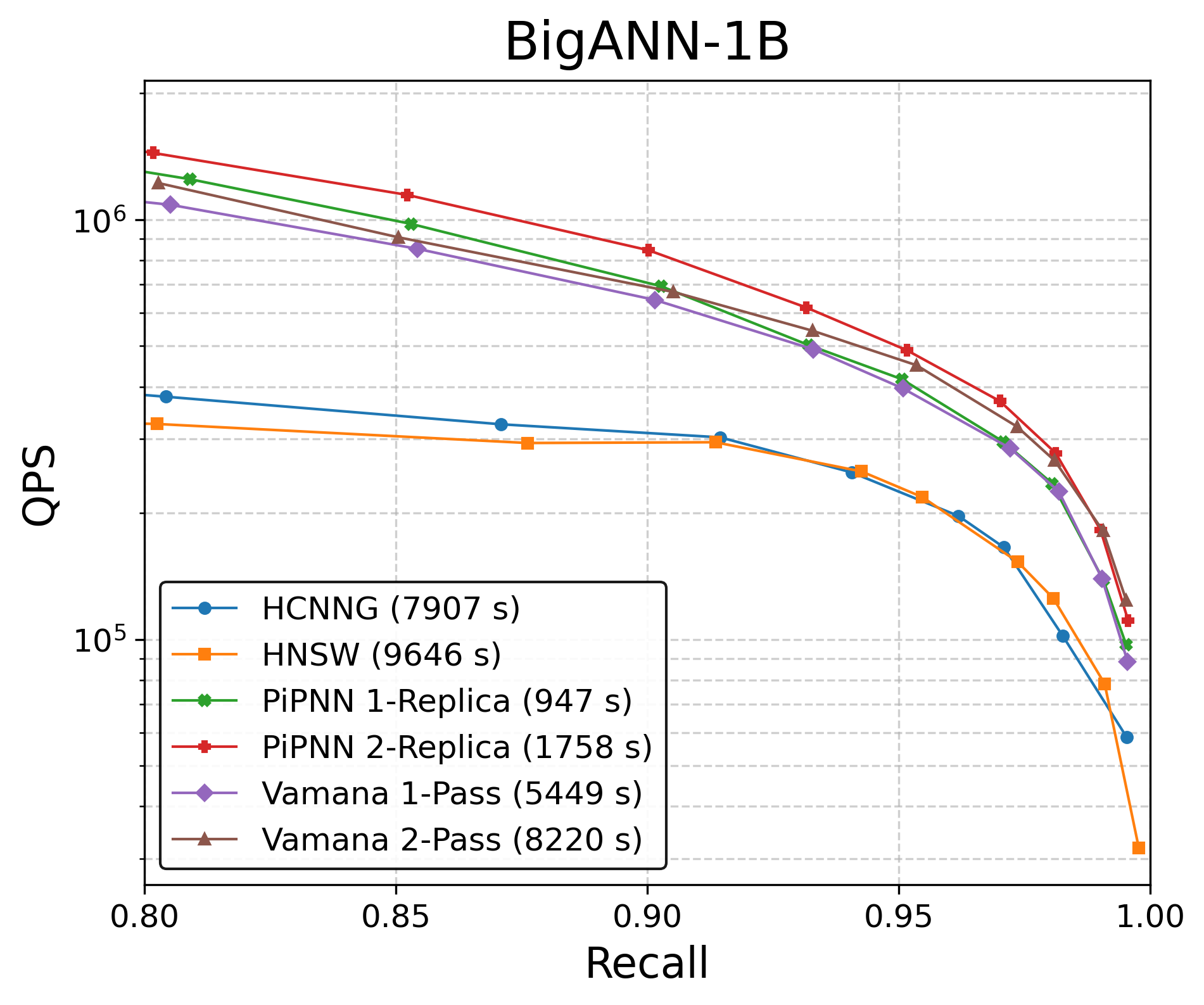}
	\end{subfigure}
	\hfil
	\begin{subfigure}[b]{.3\textwidth}
		\centering
		\includegraphics[width=\textwidth]{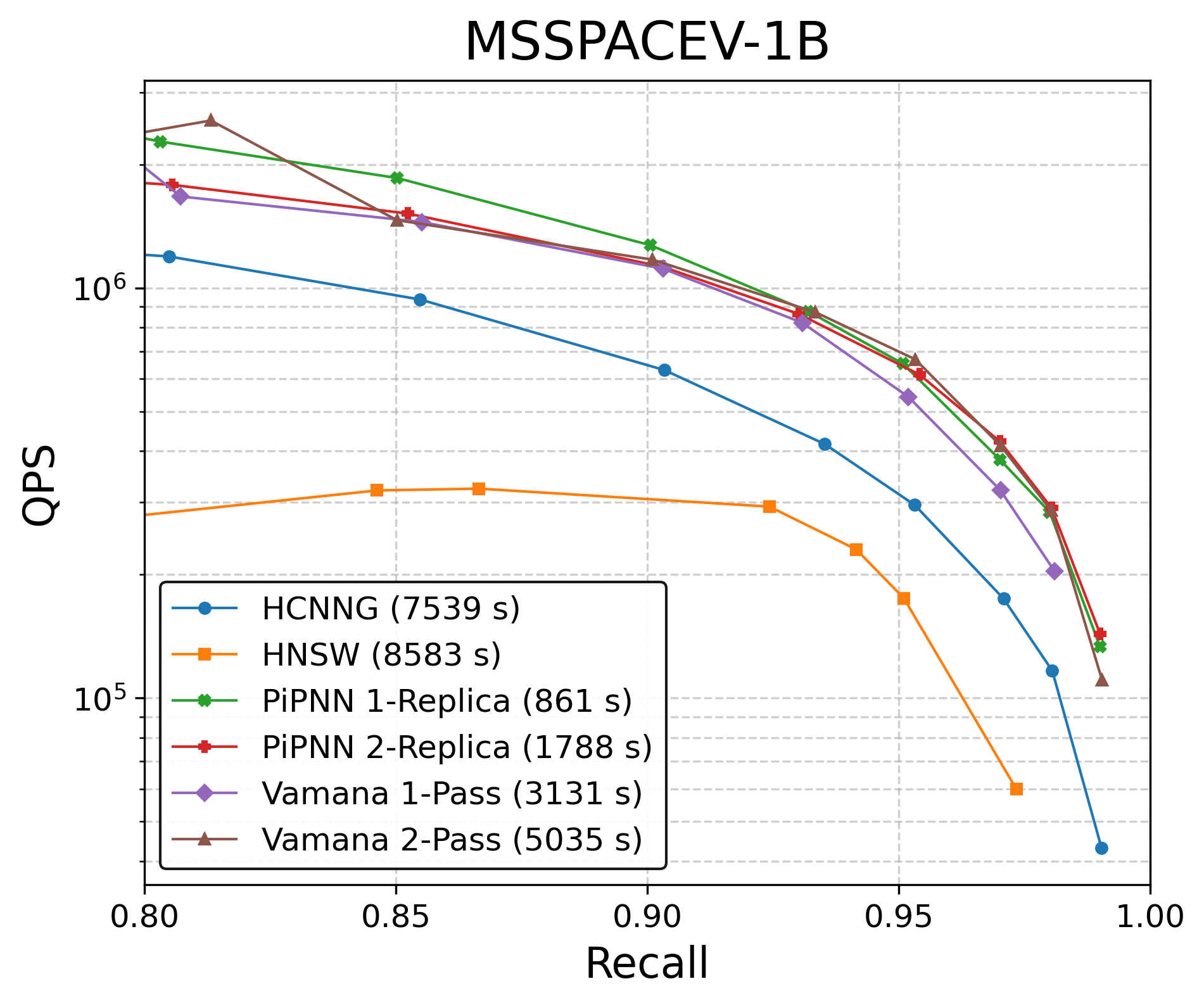}
	\end{subfigure}
	\hfil
	\begin{subfigure}[b]{.3\textwidth}
		\centering
		\includegraphics[width=\textwidth]{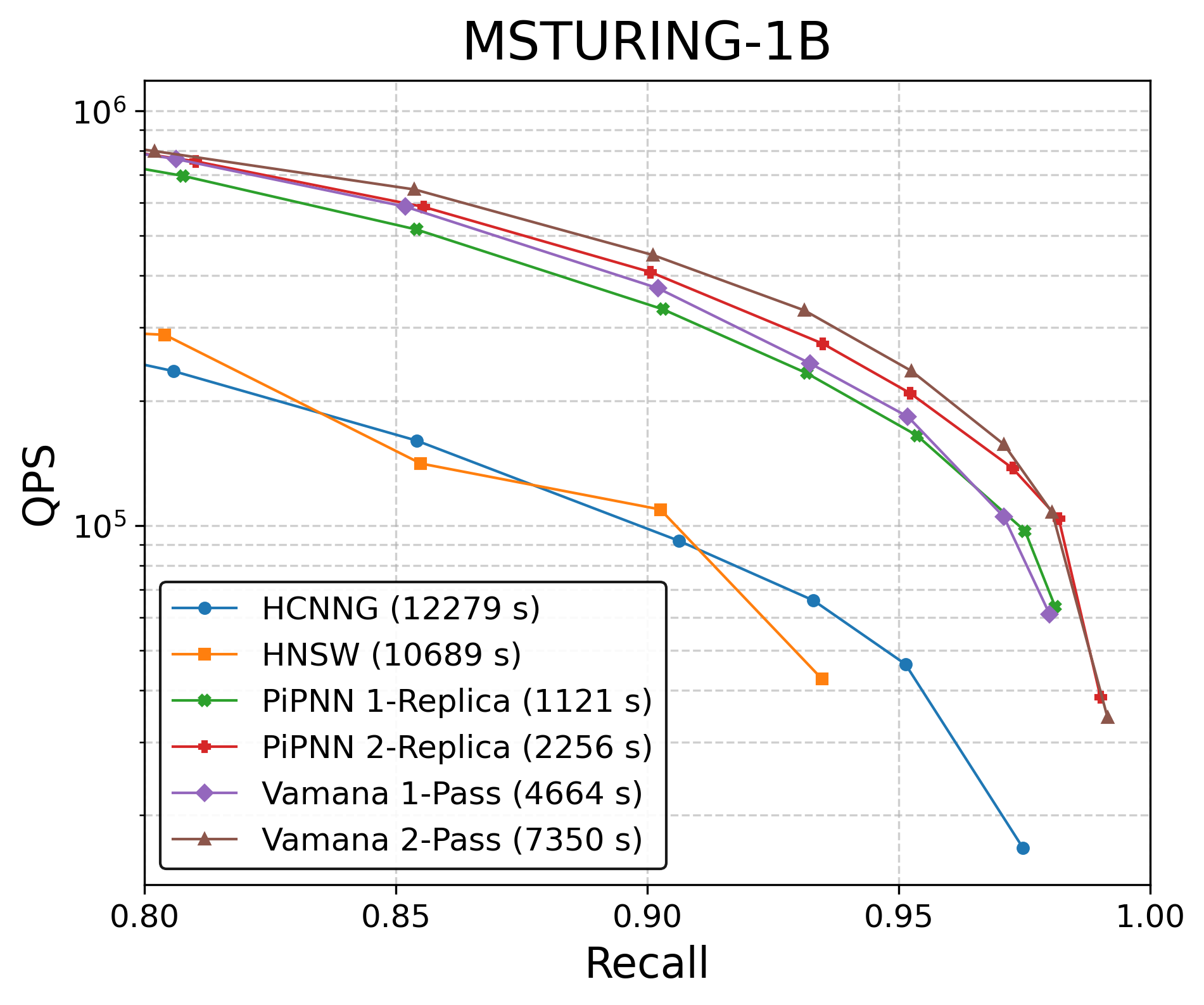}
	\end{subfigure}
    
	\begin{subfigure}[b]{.3\textwidth}
		\centering
		\includegraphics[width=\textwidth]{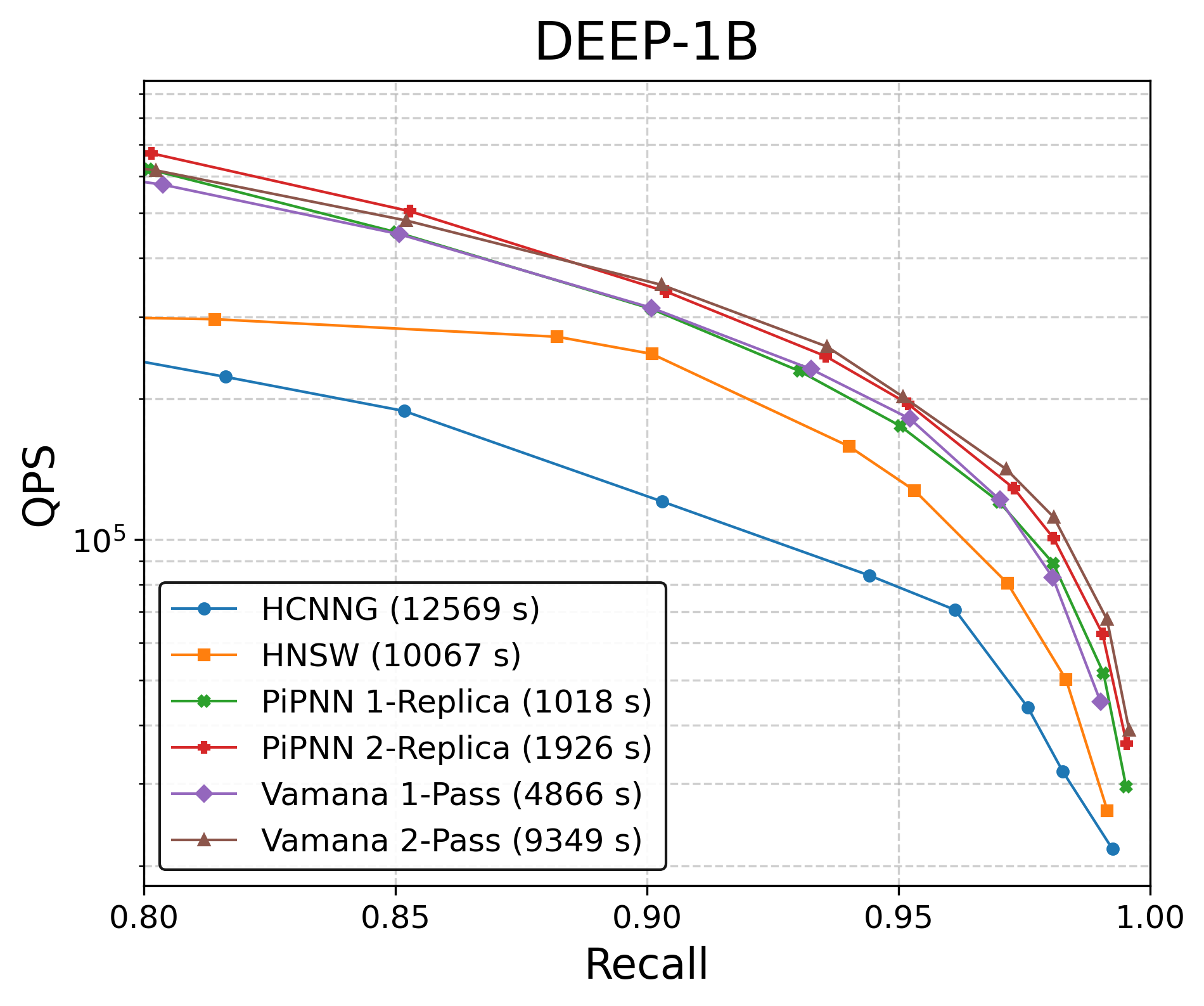}
	\end{subfigure}
	\hfil
	\begin{subfigure}[b]{.3\textwidth}
		\centering
		\includegraphics[width=\textwidth]{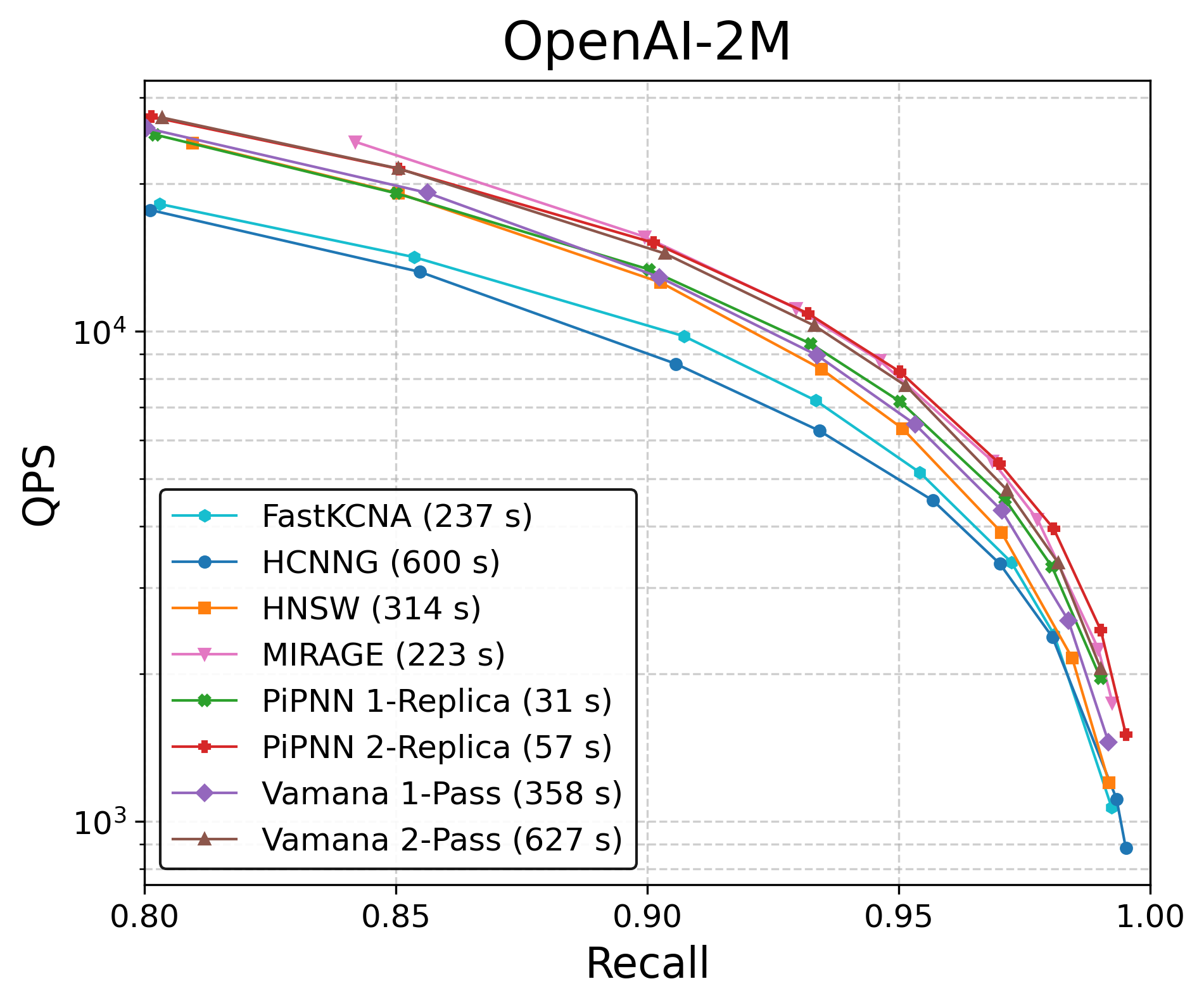}
	\end{subfigure}
    \hfil
    \begin{subfigure}[b]{.3\textwidth}
		\centering
		\includegraphics[width=\textwidth]{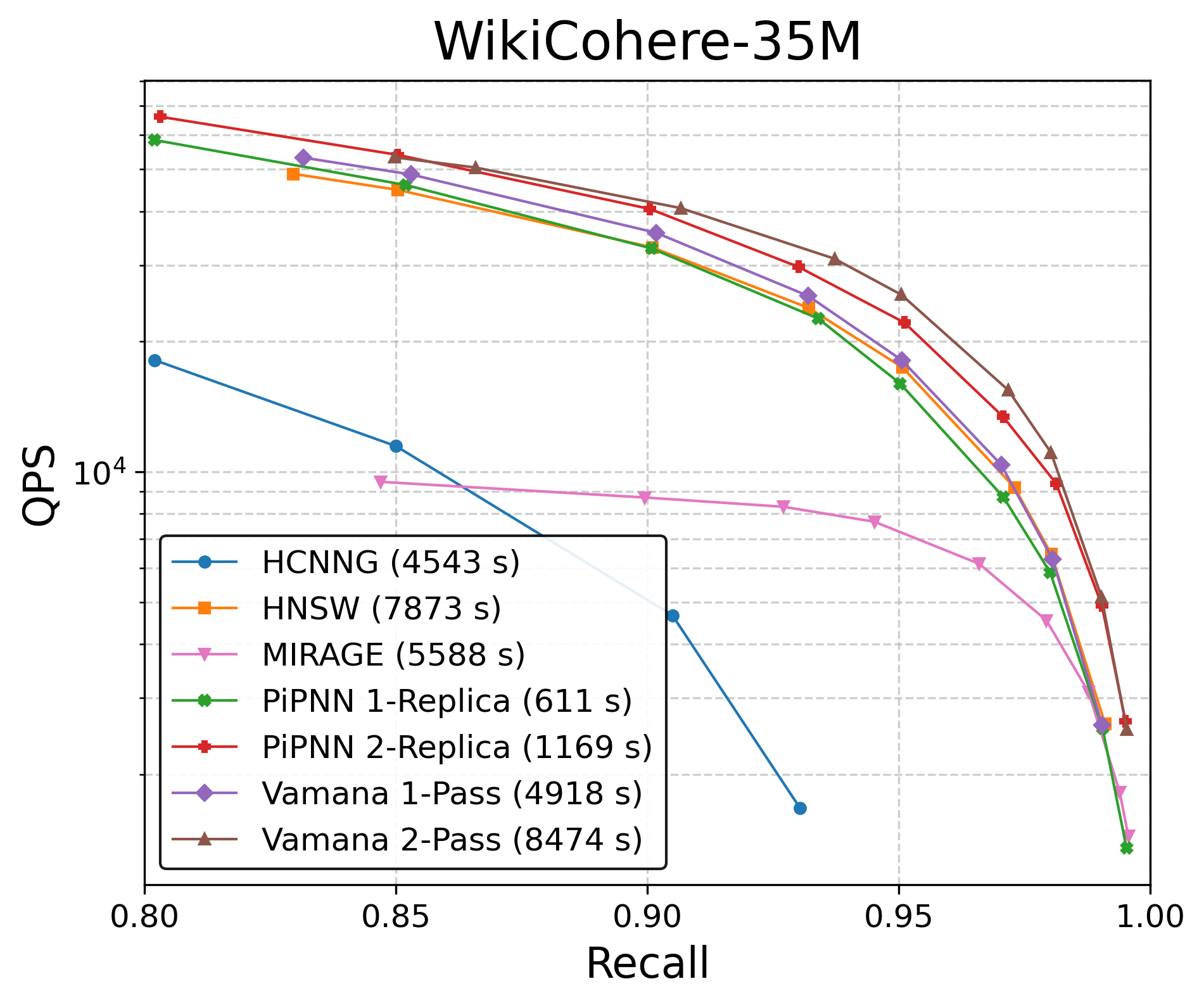}
	\end{subfigure}
    
    
	\caption{QPS vs recall for all algorithms on four billion-size datasets (BigANN-1B, MS-SPACEV-1B, MS-TURING-1B, and DEEP-1B) and two high-dimensional datasets (OpenAIArXiv-2M and WikipediaCohere-35M). Build times are listed in the legends in seconds.
}
\label{fig:algorithm_experiments}

\end{figure*}

\subsection{\ourmethod{} vs. Existing ANNS Methods}\label{subsec:comparison}


We now turn to evaluating the best instance of \ourmethod{} that we identified against state-of-the-art methods Vamana and HNSW, as well as HCNNG. 
We demonstrate that \ourmethod{} consistently produces indexes of comparable quality to the best across all datasets. We then show that \ourmethod{} always produces indexes much more quickly than rival methods, and document that as expected the superior data-locality of \ourmethod{} contributes to this result. We conclude by applying the considered ANNS methods to the downstream task of building a high quality $k-$NN graph, where \ourmethod{}'s fast construction times lead to a more than 2x speedup over HNSW.

\myparagraph{Index Quality}
The results of our experiments are shown in Figure~\ref{fig:algorithm_experiments}. The runs on OpenAI and WikiCohere show that \ourmethod with a single replica yields graph indexes of equivalent quality to Vamana with one pass. The improved performance that Vamana achieves by adding an extra pass can be matched in \ourmethod by performing an extra \emph{replicate}, i.e., an independent copy of the RBC algorithm. 
This behavior remains consistent on all of our billion-scale datasets. 
As explained in \Cref{sec:prelims}, the recall numbers reported in our experiments are specifically $10@10$ recall in accordance to existing literature, but we verify that the relative performance of the methods remains consistent at $100@100$ recall (see \Cref{sec:additional_experiments}).

Our runs with an extra replicate show that \ourmethod is able to trade additional computation to build higher quality indexes with corresponding increases in the build time.
We note that such a tradeoff is not possible for HCNNG.
In particular, as HCNNG increases the number of replicas, it unavoidably increases the average degree (and maximum degree) of the graph, increasing the number of distance comparisons and degrading query performance.
Additional replicas in HCNNG also require buffering three candidates per point, per replica, and thus linearly increases the memory usage. For a complete empirical breakdown of the peak memory footprint for \ourmethod compared to our baselines, please see \Cref{tab:ablation-build-peak-memory}.

MIRAGE was unable to run on billion-scale datasets due to running out of memory.
MIRAGE was able to run on our 100M-scale datasets, but was unable to achieve above 0.8 recall.
We note that in the MIRAGE paper, the authors only evaluated their algorithm on very small (i.e. million-scale) datasets. The authors of FastKCNA also evaluated their work on million-scale datasets, though we were able to run their work on up to 100M points without running out of memory. We found FastKCNA achieved on average $0.78\%$ query throughput across BigANN-100M and MS-SPACEV-100M as compared with \ourmethod{} and took average $18.64\times$ longer to build.
Our results show that \ourmethod{} is significantly faster than all other methods.
The highest quality indexes constructed by \ourmethod{} (2-replica \ourmethod{}) are \emph{always} built faster than 1-pass Vamana, while always being superior in quality (typically equivalent to 2-pass Vamana).

\myparagraph{Index Construction Times}
On the very high dimensional datasets OpenAI and WikiCohere, we observe that \ourmethod consistently yields significant speedups over HCNNG, HNSW, Vamana, MIRAGE, and FastKCNA. \ourmethod{} builds indexes 8-20$\times$ faster than other methods, or 4-10$\times$ faster when performing an extra replica for better query performance. On the billion-scale datasets, \ourmethod builds 4-12$\times$ faster than HCNNG and Vamana with a single replica, and 2-6$\times$ faster when doing an extra replica.

\myparagraph{Memory Accesses and Instructions}
\ourmethod{}'s performant index construction arises from better data locality leading to an order of magnitude fewer cycles and LLC-misses than Vamana and HNSW. It thus achieves superior instructions per cycle (1.26 vs. 0.44 and 0.33). Additionally, the use of SIMD yields an improvement in the number of instructions needed. 
We give a breakdown of the cycles, instructions, and last level cache misses for \ourmethod{}, Vamana, and HNSW during index construction in Supplemental Table \ref{tab:perf-averages}. 

\begin{figure}
    \centering
    \includegraphics[width=0.9\linewidth]{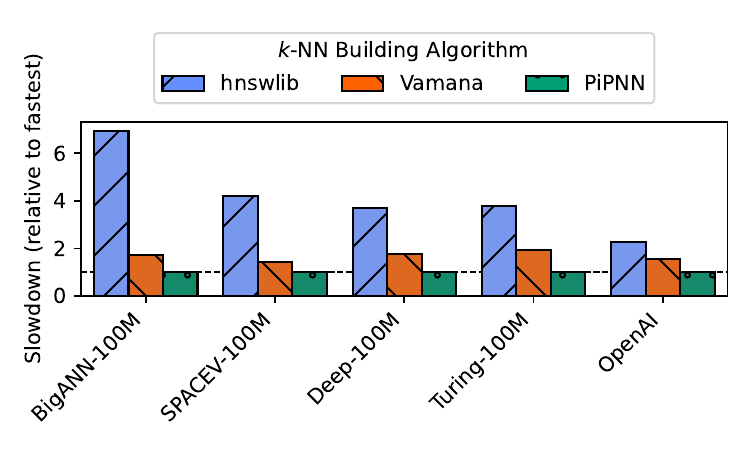}
    \caption{Performance of $k$-NN graph building for $k=10$ using different ANNS methods. The $k$-NN graphs have at least 95\% recall.
    }
    \label{fig:knnconstruction}
\end{figure}

\myparagraph{Building $k$-NN Graphs}
As an application of \ourmethod{}'s faster index construction, we consider the problem of constructing a $k$-NN graph for $k=10$ that has at least 95\% recall for the $k$-NN edges.
We note that building a high-quality $k$-NN graph is a key substep of numerous  unsupervised clustering algorithms~\cite{yu2025dynhac,shi2021scalable,dhulipala2021hierarchical,dhulipala2023terahac}, as well as large-scale data de-duplication pipelines~\cite{gu2025streaming,carey2022stars}.
We performed a grid search over parameters used to tune index quality and search quality for each algorithm to minimize the total time to construct the final $k$-NN graph.

Figure~\ref{fig:knnconstruction} shows the results of our $k$-NN construction, showing the overall slowdown of each method over \ourmethod{}, which is always the fastest.
Although \ourmethod{} yields much larger speedups if one simply measures the index build, in this application, this task requires not only building an index, but to query it for all points, and the time for this latter step is roughly the same for all high-quality indexing methods.
We find that compared to hnswlib's HNSW implementation \ourmethod{} is between $2.2$--$6.9$x faster.
Compared to ParlayANN's highly optimized Vamana implementation, \ourmethod{} is between $1.4$--$1.7$x faster in its end-to-end time.

\section{Discussion and Conclusion}
With \ourmethod{}, we have devised a novel partitioning-based algorithm for constructing graph-based nearest neighbor search indexes which builds indexes up to 12.9x faster than the prior state of the art. 
This enables significant speedups for downstream tasks, as shown in our application experiment, wherein we achieve up to 6.9x speedups over HNSW for high quality $k-$NN construction. Moreover, we have shown in our ablation study that the general framework of \ourmethod{} admits a rich design space, and multiple strategies can be employed to design a family of competitive algorithms. Our \HashPrune{} algorithm makes online pruning extremely easy and, by guaranteeing a fixed degree, allows us to trade more computation for better index quality which results in index quality on-par with 2-pass Vamana.

Our results open several promising lines for future study. Firstly, it remains to be shown what further benefits we could get by use of quantized GEMM operations (e.g., on scalar-quantized points). 
Secondly, since the basic operation of \ourmethod{} is matrix multiplication, we could make use of hardware acceleration from GPUs or TPUs to further accelerate index construction. 
Moreover, while \HashPrune{} alone yields high quality indexes, it remains to be seen whether the method can be augmented to achieve its best performance without the use of a final prune.
Lastly, we believe that \ourmethod{}'s approach is a natural fit for distributed data processing, and are interested in studying how \ourmethod{} can scale when building graphs on very large scale inputs (e.g., tens of billions of points).

\section{Acknowledgement}
This work is supported by NSF grants CCF-2403235 and CNS2317194 and by the National Science Foundation Graduate Research Fellowship Program under Grant No. DGE 2236417. Any opinions, findings, and conclusions or recommendations expressed in this material are those of the author(s) and do not necessarily reflect the views of the National Science Foundation.


\bibliographystyle{ACM-Reference-Format}
\bibliography{references}

\clearpage
\appendix
\section{Appendix}
\subsection{Partitioning Methods}\label{subsec:partitioning_ablation}
In this section, we study several different methods for generating overlapping partitions of the point-sets 
including: (1) random binary partitioning; (2) the RBC method described in the main text; (3) hierarchical k-means; and (4) sorting-LSH.
We then experimentally evaluate our choices for partitioning across different datasets and show that RBC is a high-quality and scalable default partitioning method for \ourmethod{}.

\subsubsection{Binary Partitioning}
Given a subproblem $\mathcal{P} \subseteq \mathcal{X}$, \emph{binary partitioning} selects two distinct leader points $p_1, p_2 \in \mathcal{P}$ at random, then dividing $\mathcal{P}$ into two parts $\mathcal{P}_1 = \{p \in \mathcal{P} : \lVert p, p_1 \rVert \leq \lVert p, p_2 \rVert\}$ and $\mathcal{P}_2 = \{p \in \mathcal{P} : \lVert p, p_1 \rVert > \lVert p, p_2 \rVert\}$ \cite{munoz2019hierarchical}. 
In plain words, we assign each point to its closest leader. 
The two resulting sets define the points in the subsequent subproblems, which are divided recursively by applying binary partitioning. 
This repeats until the number of points in every subproblem is at most a predefined cluster size \clustersize. 
The resulting clusters, $\mathcal{B}_1, \dots, \mathcal{B}_b$, we refer to as leaves.

Binary partitioning is the only technique used in the HCNNG algorithm; after dividing all points into leaf clusters, HCNNG proceeds by building subgraphs on every leaf, then taking the union of these subgraphs. 
Because the leaves resulting from a single partitioning procedure are disjoint, the subgraphs built on them will naturally be disconnected from each other. 
Thus, the algorithm needs some way to connect leaf clusters together. 
HCNNG resolves this through the replication process described earlier in this paper. That is, it runs the partitioning and leaf-building phases many times, using different seeds for randomization in order to produce different leaf clusters, then takes the union of all resulting graphs. 
This connects the graph because, across many replications, an individual point will share leaves with many different sets points.

While replication can be effective for combining leaves into a unified graph, it is expensive. HCNNG often requires upwards of 30 replications to produce graphs of acceptable quality. Notably, binary partitioning does not accommodate the fanout strategy in lieu of replication.

\subsubsection{Hierarchical $k$-Means}
In the well-studied problem of $k$-means clustering, the problem is to select a subset of $k$ centers from $\mathbb{R}^d$ that minimizes the sum of squared distances between any point and its nearest center. 
Although $k$-means clustering is often difficult in high-dimensional spaces, there are many heuristic solutions that work well in practice.
We implement a partitioning technique based on {\bf \emph{hierarchical $k$-means}} by applying \Cref{alg:Partition}, but rather than selecting random leaders in each subproblem, we instead run a $k$-Means algorithm to select $k$ \emph{cost-minimizing leaders} which we then use for ball-carving. 
The remainder of the algorithm is the same, including the fanout procedure. We find that choosing leaders through $k$-means yields a graph of similar (but slightly worse) quality to choosing leaders at random (see Figure~\ref{fig:partitioning_methods} in Section~\ref{sec:exps}).

\subsubsection{Sorting LSH}
Locality-Sensitive Hashing (LSH) is a celebrated technique for clustering points for the purpose of nearest neighbor search~\cite{andoni2015practical, pham2022falconn, charikar2002similarity}. 
A LSH family $\mathcal{H}$ is a family of hash functions of the form $h:\mathcal{X}\rightarrow\{0,1\}$, such that similar points are more likely to collide; namely, the probability $Pr_{h\sim\mathcal{H}}[h(x)=h(y)]$ should be large when $x,y\in\mathcal{X}$ are similar, and small when $x, y$ are farther apart. 
Following~\cite{gottesburen2024unleashing}, we create a {\bf \emph{Sorting LSH}} index by hashing each point $x\in R$ multiple times via independent hash functions $h_{1},...,h_{t}$ from $\mathcal{H}$, concatenating the hashes into a string $(h_{1}(x),h_{2}(x),...,h_{t}(x))$, and then sorting the points in $R$ lexicographically based on these strings of hashes.
We then partition into leaves by dividing the sorted sequence into consecutive groups of at most $mx$ points.

\begin{figure}
    \centering
    \includegraphics[width=\linewidth]
    {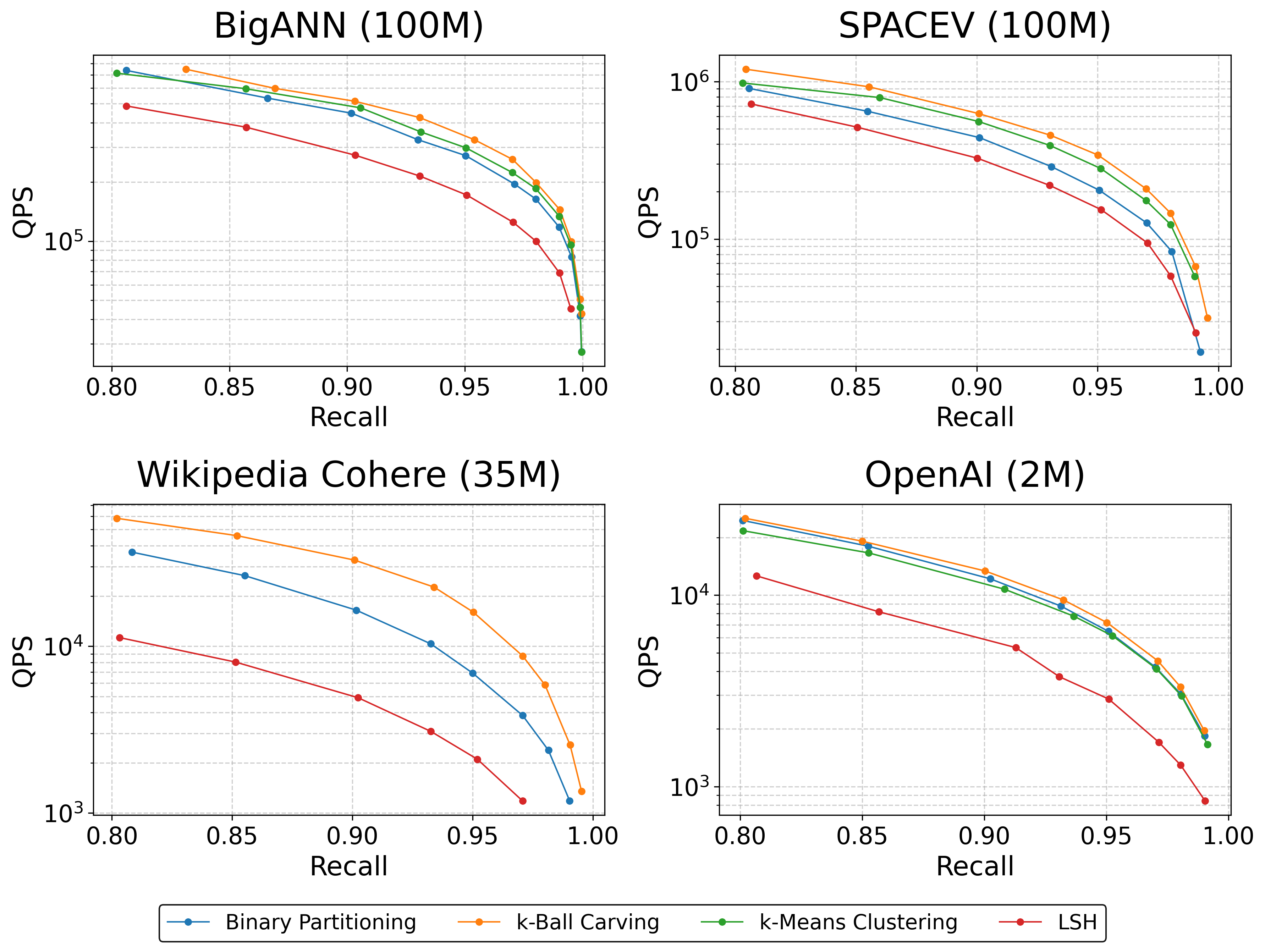}
    \caption{Four different partitioning methods applied in the \ourmethod algorithm to different datasets. All four methods use a 2NN-graph in the leaf-building phase. $k$-means is omitted from the WikiCohere plot due to failure to finish within 24h.}
    \label{fig:partitioning_methods}
\end{figure}

Since there is no natural analog of fanout for a sorting LSH index, we rely on replication instead of fanout to place a point in multiple leaves.
Although constructing a sorting LSH index is quite fast (it requires applying a number of hyper-plane tests per-point, and some parallel sorting), the extra cost of performing replication when compared to multi-level fanout offsets much of the faster computation of LSH-regions compared to dense-ball carving, to the point of actually being slower. 
We further find that LSH-based partitioning yields a significantly lower quality graph than $k$-ball carving, either with random leaders or $k$-means leaders (Figure \ref{fig:partitioning_methods}).

\myparagraph{Comparison of partitioning strategies}
We have now described four partitioning methods for use in \ourmethod: binary partitioning, dense-ball carving with either random leaders or $k$-means centroids, and Sorting-LSH. To demonstrate the merits of the partitioning methods in isolation, we fix the leaf strategy to the best performing method (bidirected $k$-NN).
The relative performance between partitioning strategies was similar when using other leaf methods.

\Cref{fig:partitioning_methods} shows the QPS-recall tradeoff of the different partitioning methods on the ablation datasets. 
Randomized ball carving consistently produces the best query performance. 
Binary partitioning and $k$-means are sometimes competitive (OpenAI, BigANN). Table \ref{tab:partitioning_times} shows the time taken by the different partitioning methods. 
Randomized ball carving is consistently much faster than $k$-means and Sorting-LSH. 
Compared to binary partitioning it is substantially faster on OpenAI and Wiki, while exhibiting similar running time on BigANN and SpaceV.
We omit the results for $k$-means clustering on Wikipedia Cohere because it failed to complete within 24 hours. 
Although $k$-means could be accelerated, e.g., with subsampling strategies, as the index quality is not better than ball carving we did not optimize the indexing time further. 
In summary, randomized ball carving consistently achieves the highest quality indexes while having the fastest construction time.

\begin{table}[]
\small{}
    \centering
    \caption{Time spent on each partitioning method. Listed times are in seconds and only include the partitioning phase.}
    \label{tab:partitioning_times}
    \begin{tabular}{|l|c|c|c|c|}
        \hline
        Method & BigANN & SPACEV & OpenAI & Wiki \\
        \hline
        Binary Partitioning         & 210.3 & 183.4 & 99.3 & 971.2 \\
        Rand. Ball Carving     & 224.3 & 220.7 & 10.2 & 417.9 \\
        $k$-Means Clustering        & 896.6 & 856.4 & 110.0 & N/A \\
        Sorting LSH                 & 469.5 & 369.9 & 110.7 & 1212.3 \\
        \hline
    \end{tabular}
\end{table}

\subsubsection{Fanout vs. Replicas}\label{sec:fanout_ablation}

\begin{figure}
    \centering
    \vspace{-1em}
    \includegraphics[width=\linewidth]{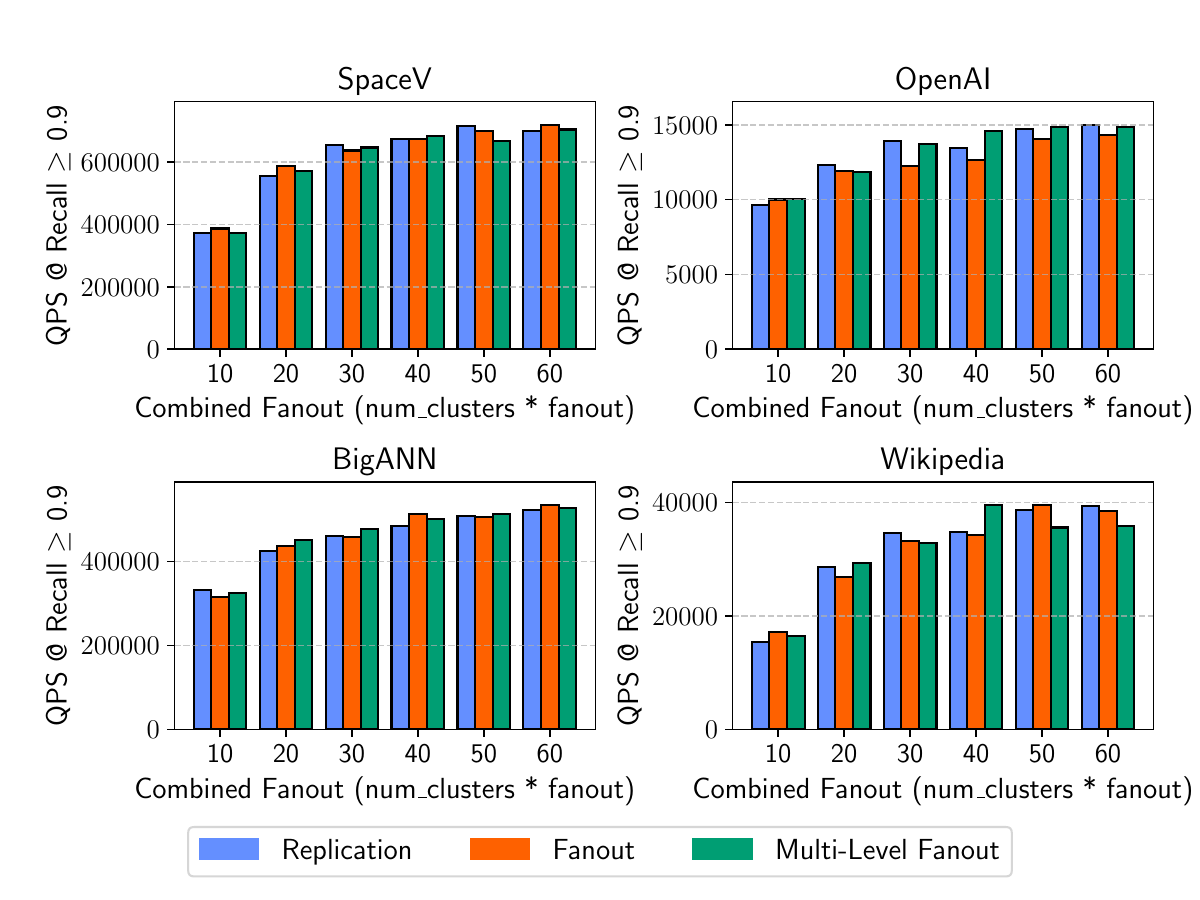}
    \vspace{-1em}
    \caption{Given equal point repeat budget, all three methods exhibit similar QPS at fixed recall. 
    }
    \label{fig:qps_replication_bar}
\end{figure}

In \Cref{sec:partitioning}, we explain that \ourmethod{}'s RBC method uses a single recursive process to select multiple leaders per level, referred to as multi-level fanout, instead of repeatedly replicating the entire partitioning. We showed that this significantly speeds up partitioning times \Cref{fig:build_replication_bar} and had no impact on quality. We show the lack of quality impact in \Cref{fig:qps_replication_bar}. As can be seen, the difference in QPS at $0.9$ recall is marginal and without any consistent winner. Any differences in the results can likely be explained by OS jitter.

\subsection{Fanout vs. Replication}\label{sec:fanout_vs_replication}

In \Cref{sec:partitioning}, we presented fanout and multi-level fanout as replacements to replication for strategies to create overlapping partitions. Here, we give a more thorough explanation for why fanout and multi-level fanout yield significantly better build times.

Suppose we select $\ell$ leaders in each subproblem during partitioning. Then within a given level of the recursion tree, we must obtain the distances between each point and the leaders within the subproblems in which it is contained. In other words, we compute a total of $\AggrFan \cdot \lvert \mP \rvert \cdot \ell$ distances in the $i$th level, where $f_i$ is the aggregate fanout (i.e. how many subproblems each point is contained in) at level $i$. Under the replication strategy, $\AggrFan = 1$ at every level, but we repeat the entire procedure some $r$ many times, yielding the same cost as if each point had aggregate fanout $r$ at every level. If we were to get an analogous partitioning under the fanout strategy, we would execute the top level only once, wherein we would assign each point to the subproblems of its $r$ nearest leaders. Levels beyond the first would have $f_i = r$ and thus have the same number of distance computations as with replication, but the first level wpuld only call for $\lvert \mP \rvert \cdot \ell$. Multi-level fanout extends the cost reduction beyond the first level. For example, if we fanout $r_1$ on the top level and $r_2$ on the second level, then the second level now maintains a low aggregate fanout of $r_1$.

In summary, the strategies require the following number of distance computations:\\
\hspace*{1em} $r$ replications\\
    \hspace*{3em}$r \cdot \lvert \mP \rvert \cdot \ell$ each level\\
\hspace*{1em} $r$ fanout\\
    \hspace*{3em}$\lvert \mP \rvert \cdot \ell$ first level, $r \cdot \lvert \mP \rvert \cdot \ell$ thereafter\\
\hspace*{1em} $r_1, r_2$ multi-level fanout (where $r_1 \cdot r_2 = r$)\\
    \hspace*{3em}$\lvert \mP \rvert \cdot \ell$ first level, $r_1 \cdot \lvert \mP \rvert \cdot \ell$ second level, $r \cdot \lvert \mP \rvert \cdot \ell$ thereafter
\noindent As a reminder, the recursion depth is usually very low because of the high number of leaders selected in each subproblem. Thus, reducing the cost of the top 2 levels of recursion has a very large impact on overall partitioning time. This also limits how many levels we can distribute fanout across, as any fanout beyond the second level has a high likelihood to not take effect due to many subproblems already reaching their base cases.

\subsection{Picking within Leaves}\label{sec:picking_ablation}

\begin{figure}
    \centering
    \includegraphics[width=\linewidth]{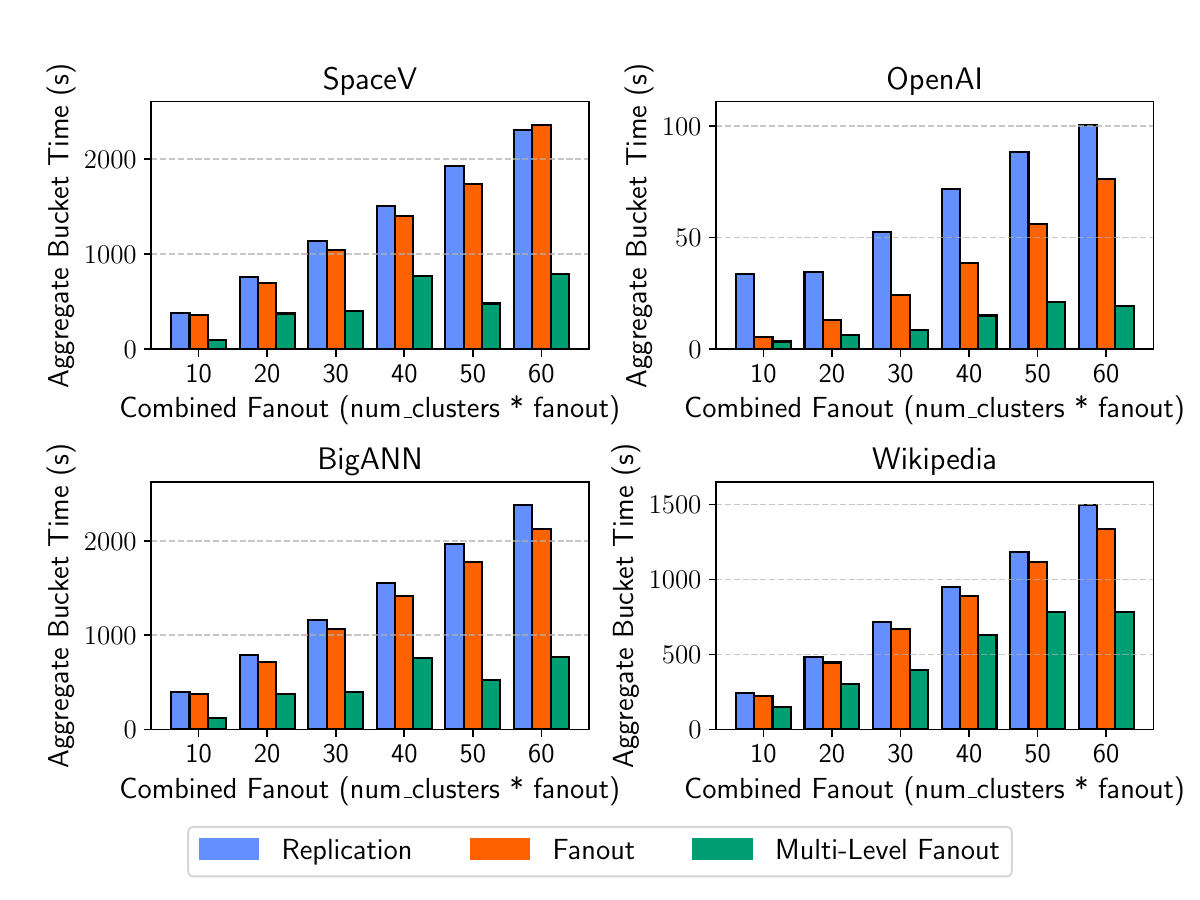}
    \caption{Given equal point repeat budget, Multi-level Fanout greatly reduces time to partition into leaves (termed `Buckets' here).}
    \label{fig:bucket_replication_bar}
\end{figure}

As mentioned, \ourmethod{} selects for each element of a leaf a subset of its co-inhabitants to prune via \HashPrune. In our final algorithm we used bi-directed k-NN's for this purpose. In this section we describe all the other methods we considered, and give an ablation to demonstrate why k-NN was suitable for inclusion as the sole method used in \ourmethod{}.

\subsubsection{Degree-Restricted Minimum Spanning Trees}
We first describe an existing leaf method developed for use in the HCNNG algorithm that we incorporate into \ourmethod{}.
The method is to build an undirected degree-restricted minimum spanning tree over the points within a leaf, where the degree of each node is capped to be at most $3$. We will refer to this approach simply as \textbf{MST}.
This approach enforces connectivity on the subgraph while (1) prioritizing the shortest edges and (2) ensuring that the resulting graph is sparse.
Intuitively, these short edges are important for the final stages of a greedy search, as they increase the likelihood of finding the true nearest neighbor once the search has navigated to the correct region of the graph.

However, building an MST using all pairwise distances within a leaf is a performance bottleneck due to the cost of storing and sorting all pairwise edges (required e.g., when running a degree-restricted variant of Kruskal's algorithm). 
To solve this, the implementation of HCNNG in ParlayANN builds the MST from a sparser graph where each point is only connected to its $l$-nearest neighbors (e.g., $l=10$) within the leaf.
This restriction was shown not to affect the quality of the resulting index, and yields significant performance improvements.

\subsubsection{$k$-Nearest Neighbor Graphs}
Extrapolating from the idea that short-range edges are important for traversing small local neighborhoods, a very natural subgraph to build on a leaf is a $k$-nearest neighbor ($k$-NN) graph. 
That is, for each point $v$, we identify the $k$ nearest points within the cluster and add edges to and/or from each of those points. 
The resulting $k$-NN graph is closely related to an MST but with a few distinctions. 
Most notably, in contrast to the MST, the $k$-NN graph on a leaf is not necessarily connected. 
However, it's not clear that this property is necessary, given that leaves themselves are not connected to each other until replication/fanout adds overlapping subgraphs. 
In exchange, the $k$-NN graph is faster to compute and more tunable.

\myparagraph{k-NN Leaf Design Space}
We consider several choices for directing the edges in the $k$-NN graph. 
In the standard notion of a $k$-NN graph, which we will refer to as the {\bf Directed \textit{k}-NN}, we add an edge from each point to its $k$ nearest neighbors. 
Alternatively, we can flip the directions of all edges in the {\bf Inverted \textit{k}-NN}, in which we add an edge \emph{to} each point \emph{from} its $k$ nearest neighbors. 
Although unconventional, this inverted $k$-NN construction may in fact be better suited to building a navigable graph than the standard $k$-NN graph. 
This is because if a greedy traversal for a point in the dataset lands on its nearest neighbor, the only node to which it can move that makes progress to the target point is that point itself. 
More generally, we can describe this relationship as the following: if a traversal aims to reach a node, the nodes that the traversal is most likely to try to reach it from is its nearest neighbors. 
Finally, we consider the {\bf Bi-directed \textit{k}-NN}, the union of the directed and inverted $k$-NNs. In this graph, for each point, we add edges to and from its $k$ nearest neighbors. We will also refer to {\bf \textit{k}-NN} as shorthand for this method, as we find it to be the best of the three.

Our experiments (\Cref{fig:leaf_qps}) show that the inverted $k$-NN strategy yields graphs of much higher quality than the directed $k$-NN, but that the bi-directed $k$-NN is better than both. 
This result suggests that the forward edges and backward edges each contribute to the traversability of the graph. 
Moreover, we find that building bi-directed $k$-NN on leaves with the right choice of $k$
yields better search performance than building MSTs.
Through ablations, we reveal a clear relationship between the parameter $k$ and search performance. 
Increasing $k$, as one might expect, increases the average degree of each node in the graph. 
This makes visiting a node more costly. 
In exchange, it decreases the average number of nodes visited per query. 
As we show in \Cref{fig:knn_parameter}, this results in a sweet spot for $k$ ranging from 2 to 4 which produces the best graphs.

\subsubsection{All-to-All Prune}
As described in Section~\ref{sec:incremental_graph_indexes}, the \RobustPrune procedure takes a set $\mathcal{N}$ of candidate neighbors for a point $x \in \mathcal{X}$ and returns a subset of $\mathcal{N}$ under some given size limit \cite{jayaram2019diskann}. In particular, \RobustPrune attempts to choose a set of neighbors for $x$ from the set of candidates that ensures good navigability. 
For the candidate neighbors of a point $x$, Vamana uses all points visited during a beam search for $x$. 
Typically, the vast majority of points visited during beam search are in the close vicinity of the target point. 
Thus, these points can be argued to roughly approximate a subset of the local neighborhood around $x$. 
Since the leaves into which a given $x$ falls in \ourmethod{} also approximate the local neighborhood around $x$, we consider using \RobustPrune to build leaves. 
More specifically, for a leaf $\mathcal{B}_i$, we run \RobustPrune on each $x \in \mathcal{B}_i$, supplying $\mathcal{B}_i$ as the candidate set. This yields up to \mstdeg neighbors per point in the leaf.

\myparagraph{Optimizing RobustPrune}
We note that \RobustPrune can be costly to run on large lists of candidates, because each time a neighbor is chosen, it considers all other candidates for removal.  
In \ourmethod{}, we provide \RobustPrune with several times more candidates than Vamana often does (our cluster size $C$ is roughly 1000).
To remedy this, we use a lazy implementation of \RobustPrune. 
This version does not remove any redundant candidates upon adding a neighbor. 
Instead, prior to adding a candidate $c$ to the neighbor list, we check all previously-admitted neighbors to see if any should have pruned away $c$, then delete $c$ if so. 
Compared to the standard version, our version can potentially omit unnecessary comparisons. 
Specifically, of the comparisons that standard \RobustPrune does, it only performs those on candidates closer to the target than the last neighbor added to the output neighbor list.

\begin{figure}
    \centering
    \vspace{-1.5em}
    \includegraphics[width=\linewidth]{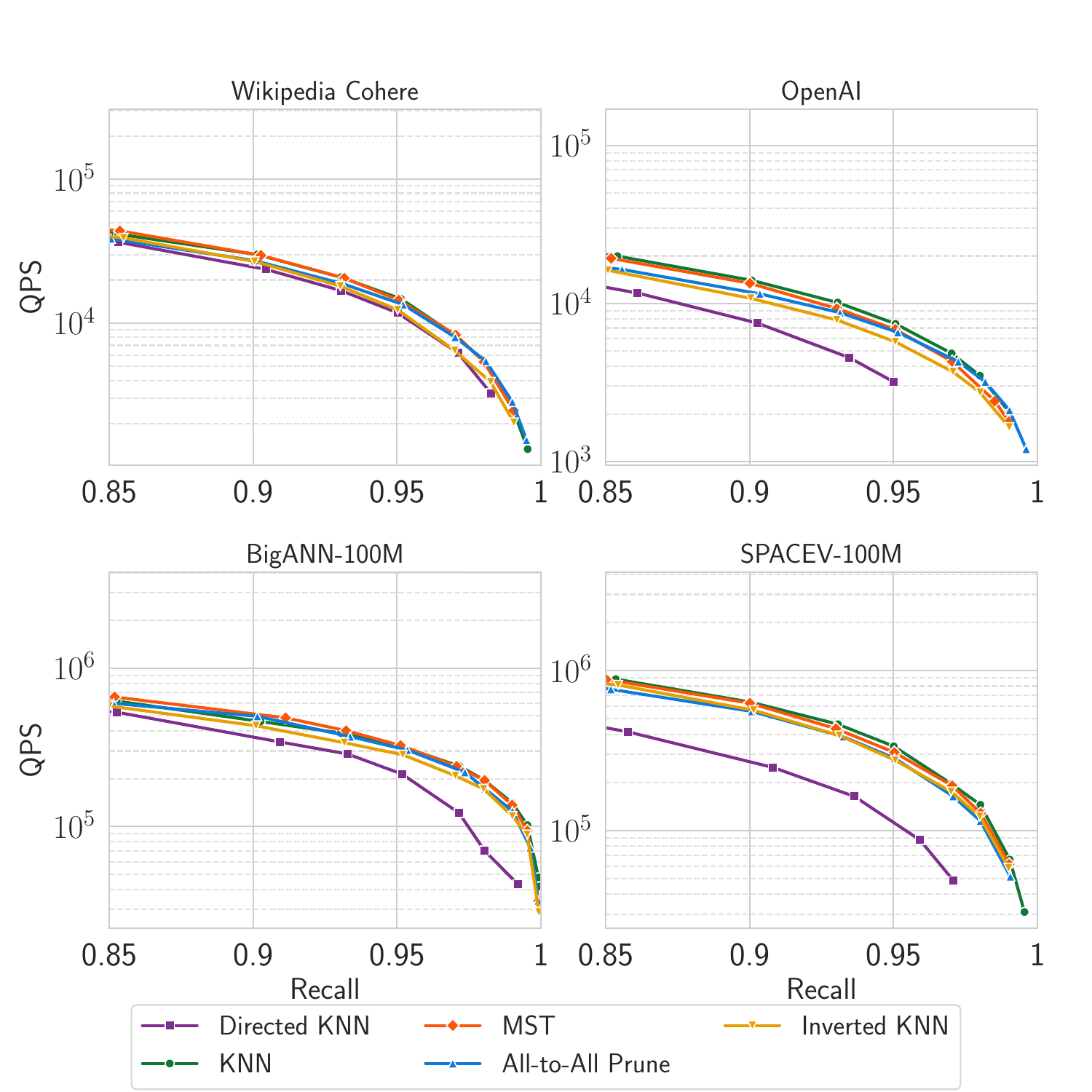}
    \caption{
   Comparing QPS-recall trade-offs of leaf pruning methods. Bi-directed $k$-NN ($k$-NN) performs best. 
   }
    \label{fig:leaf_qps}
\end{figure}

\myparagraph{Comparison of Prune Kernels} 
We evaluate five different methods for pruning the candidate lists induced by the partitioning process: building a bi-directed minimum spanning tree, RobustPrune, building a $k$-NN graph, building an inverted $k$-NN graph, and a bi-directed $k$-NN graph.
Figure \ref{fig:leaf_qps} shows the performance of these different leaf methods, keeping the rest of the pipeline fixed, on the ablation datasets. 

\begin{table}
    \caption{\small Average degree for each method for each ablation dataset. Degree is capped at 64 for all methods.}
    \label{tab:degtab}

\footnotesize
    \centering
\begin{tabular}{|l|c|c|c|c|c|}
\hline
Method & BigANN-100M & OpenAI & Space-100M & Wiki & Avg\\
\hline
Directed $k$-NN & 24.19 & 12.34 & 19.66 & 18.80 & 18.75 \\
$k$-NN & 41.04 & 21.51 & 34.47 & 32.02 & 32.26 \\
MST & 29.34 & 16.44 & 25.71 & 24.75 & 24.06 \\
Robust Prune & 62.54 & 43.07 & 60.88 & 55.31 & 55.45 \\
Inverted $k$-NN & 31.45 & 18.22 & 27.45 & 25.55 & 25.67\\
\hline
\end{tabular}
\end{table}
We observe that bi-directing the $k$-NN graphs is critical for obtaining good performance, but between bidirected $k$-NN and bidirected MST, performance is similar; bidirected $k$-NN outperforms bidirected MST in some high recall situations due to the tunability of its density. 
RobustPrune has the disadvantage of producing much denser graphs, as can be seen in Table \ref{tab:degtab}. 
This ends up producing indexes that require fewer visited nodes at high recall, but the increased number of comparisons when visiting a node outweighs the benefit. 
RobustPrune in the leaves produces candidate lists that are unlikely to be further sparsified by either HashPrune or the final application of RobustPrune. Thus RobustPrune produces graphs with average degree close to the maximum degree (55.45 on average). 

\begin{figure}
    \centering
    \vspace{-1em}
    \includegraphics[width=\linewidth]
    {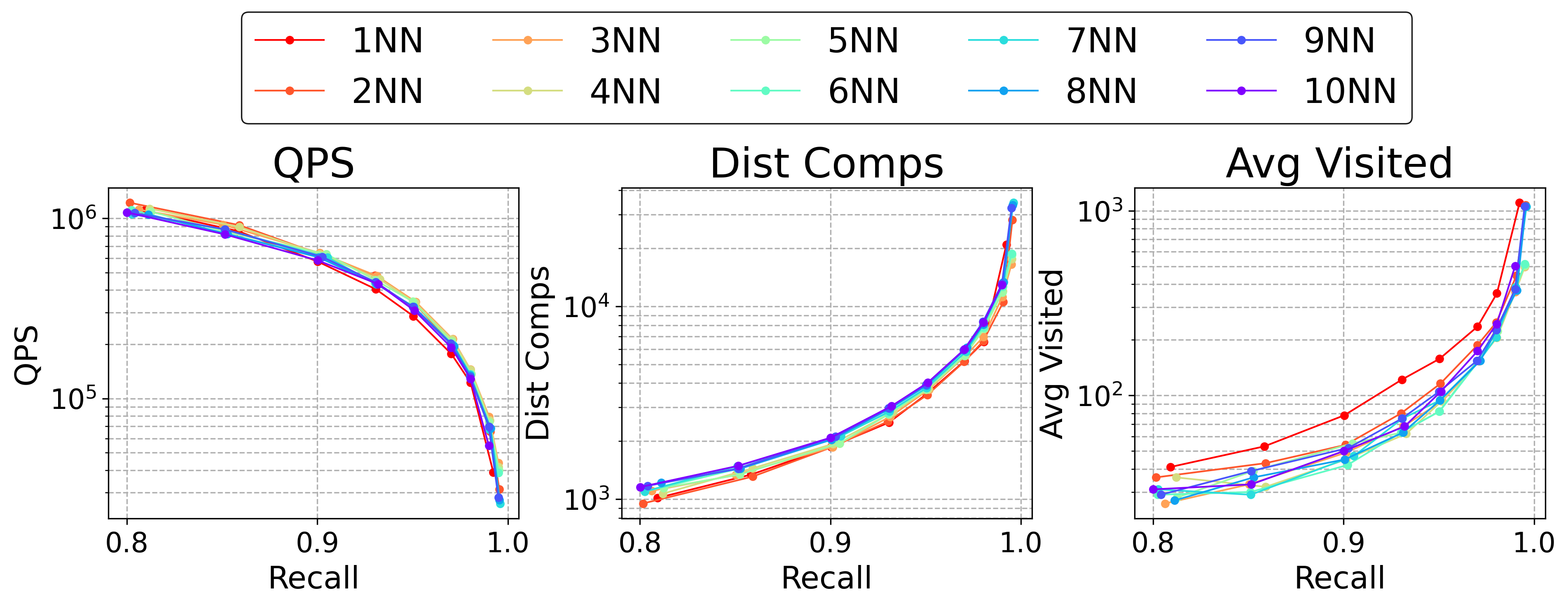}
    \caption{Varying the parameter $k$ for $k$-NN graphs in leaves.}
    \label{fig:knn_parameter}
\end{figure}

\myparagraph{Tuning $k$-NN}
By increasing the value of the parameter $k$, we can raise the density of the subgraph produced in each leaf. This increases the cost of visiting a node while decreasing the number of steps taken during a query. \Cref{fig:knn_parameter} displays this effect on MS-SPACEV across $k \in [1,10]$. The behavior is similar on other datasets. When increasing $k$, the average number of visited nodes per search goes down sharply at first but with quickly diminishing returns. The initial drop is indicative of the value of these first few short edges to the general navigability of the graph -- that is, only having an edge to one's nearest neighbor is insufficient. However, after only a few increments of $k$, the additional edges no longer clearly improve the speed at which queries converge to the target. On the other hand, the additional edges do continue to steadily increase the number of distance computations due to further densifying the graph. 

In summary, the effects of increasing $k$ from $1$ to $10$ have a significant impact on overall query performance. Increasing $k$ from $1$ to $\{2, 3\}$ yields an immediate increase to query throughput, but subsequent increases yield a steady decline in throughput due to adding redundant edges. 
Based on these results, we set $k = 2$ as a sane default, although 3 or 4 would also suffice.

\subsection{Ablation of Leaf Optimizations}\label{sec:leaf_optimization_ablation}

\begin{figure}
\vspace{-1.5em}
    \centering
    \includegraphics[width=\linewidth]{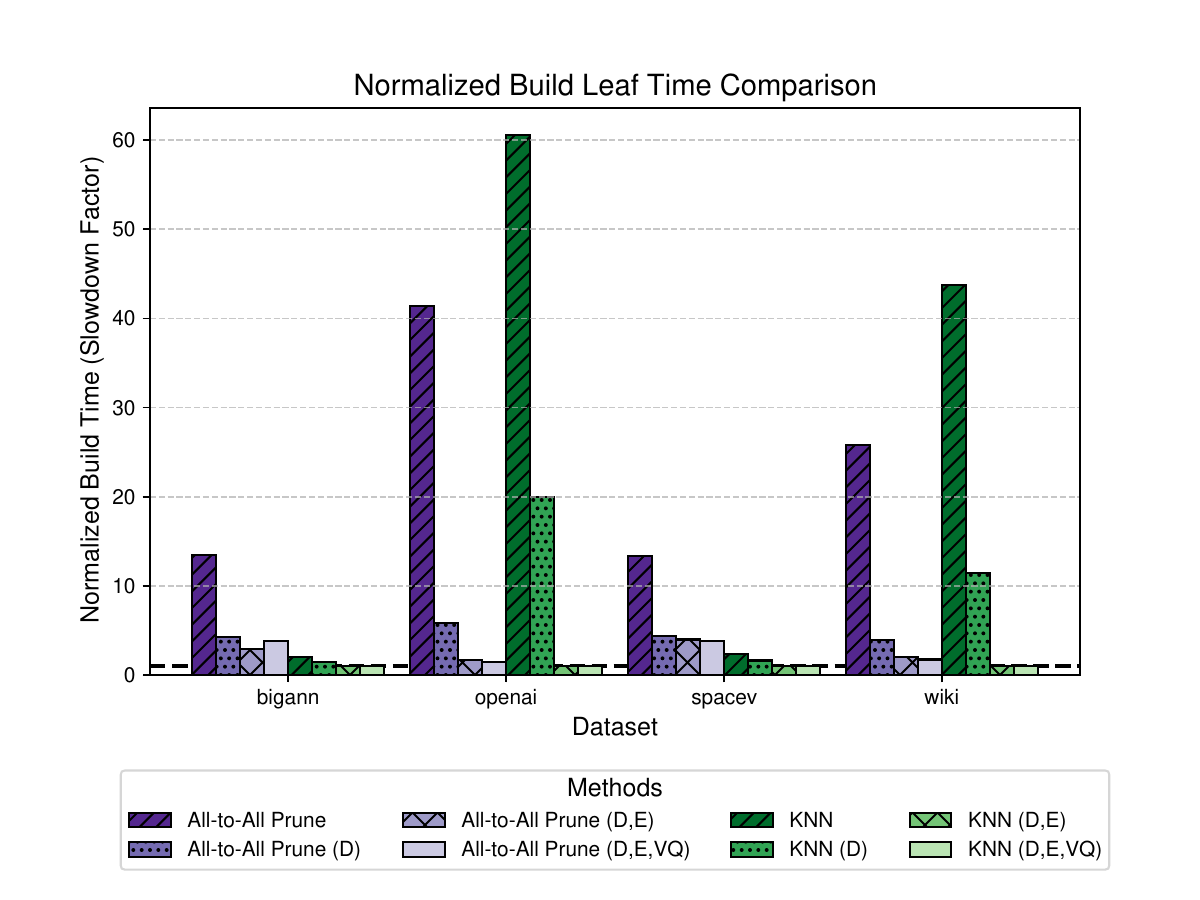}\vspace{-0.5em}
    \caption{Effect of leaf optimizations on total leaf building times. (D) denotes pre-computing the distance matrix. (E) denotes pre-computing the distance matrix with the Eigen library. (VQ) denotes using VQSort or VQPartialSort to sort distances. Data is normalized with respect to the fastest build time; lower is better.}
    \label{fig:leaf_optimization}
\end{figure}

In Section \ref{sec:pipnn_optimizations}, we described several optimizations for building leaves in \ourmethod{}, namely building distance matrices, computing those matrices using the Eigen library \cite{eigenweb}, and finding near neighbors using the Highway library's vectorized partial sort \cite{google_highway}.

We show the effect of these optimizations in Figure \ref{fig:leaf_optimization} on the ablation datasets. 
We compare naive implementations of $k$-NN as well as All-to-All Prune (wherein we perform \RobustPrune{} within each leaf to generate candidates for \HashPrune{}) against methods optimized by first building a distance matrix (D), building a distance matrix with the Eigen library (D,E), and methods which additionally use VQSort or VQPartialSort for identifying near-neighbors. 
Building a distance matrix gives average speedups of $2.44\times$ and $4.95\times$ for $k$-NN and All-to-All Prune respectively. Doing so with Eigen gives a further $8.57\times$ and $1.98\times$ speedup for the two methods. All-to-All Prune benefits marginally from using VQSort ($1.05\times$ speedup), whereas $k$-NN does not get further improvement compared to the priority queue approach. 
All optimizations combined give our final versions of $k$-NN and All-to-All Prune $27.03\times$ and $12.71\times$ speedups respectively on average.

\subsection{Intermediate and Final Pruning}\label{subsec:expprune}

\begin{figure}
    \centering
    \includegraphics[width=\linewidth]{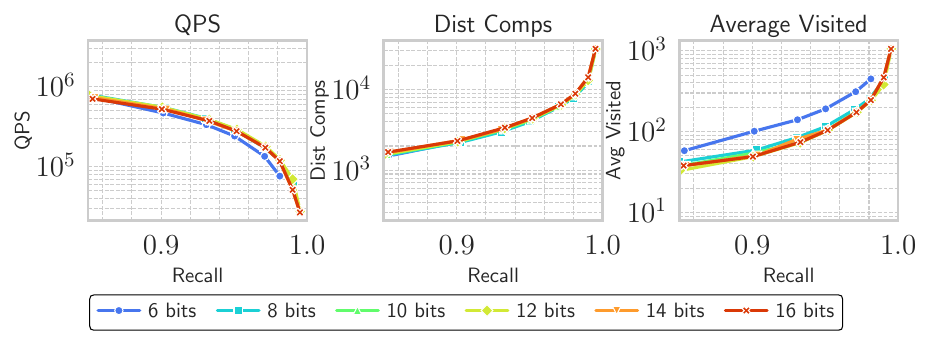}
    \caption{Parameter sweep over hash-family sizes in the range of $[6,16]$ on SPACEV (100M) dataset, without any final prune. All values between 8 and 16 are suitable.}
    \label{fig:hashbitspacev}
\end{figure}

\myparagraph{HashPrune}
We developed the HashPrune procedure to ensure that memory usage was bounded regardless of the amount of replication or the degree in a leaf. On the one hand, this is necessary to build high quality indexes on billion-scale datasets where memory requirements are high, but HashPrune also ensures that memory can be allocated up-front in a single contiguous block. Thus, the procedure not only enables better scalability, but it also avoids dynamically resizing the intermediate candidate lists. 

To test whether limiting the candidates during construction impacts query performance, we compared reservoir sizes of 64, 128, and 192 against an unbounded degree intermediate graph on the ablation datasets (with final prune performed down to 64 max degree). We observed that HashPrune has almost no impact on the distance computations performed (and, thus, the QPS achieved) to achieve high recall when paired with the bi-directed $k$-NN method we use for candidate picking candidates within buckets. These results are displayed for MS-SPACEV-100M in \cref{tab:hashbit_data}. This table also shows the marginal variation across different $m$ parameters for the hash functions. This surprising parameter insensitivity is likely caused by our use of a final prune.  

\myparagraph{Final Prune}
\ourmethod{} employs a final RobustPrune on each candidate list after the partitions have been processed. 
We found this to improve QPS by an average of $10.18\%$ and $9.45\%$ at 0.9 and 0.95 recall respectively across the four ablation datasets. 
We find that the final pruning step comprises an average of $9.82\%$ of the total build time across the four ablation datasets. This number drops to just $6.86\%$ for the 100M sized ablation datasets, where the other indexing steps are more costly.
Due to its quality improvements, and given its relatively modest impact to the overall construction time, we find that final pruning is beneficial and always enable it.

\subsection{\ourmethod{} vs. Recent ANNS Methods}\label{subsec:supp_comparison}

\begin{figure}
    \centering
    \includegraphics[width=\linewidth]{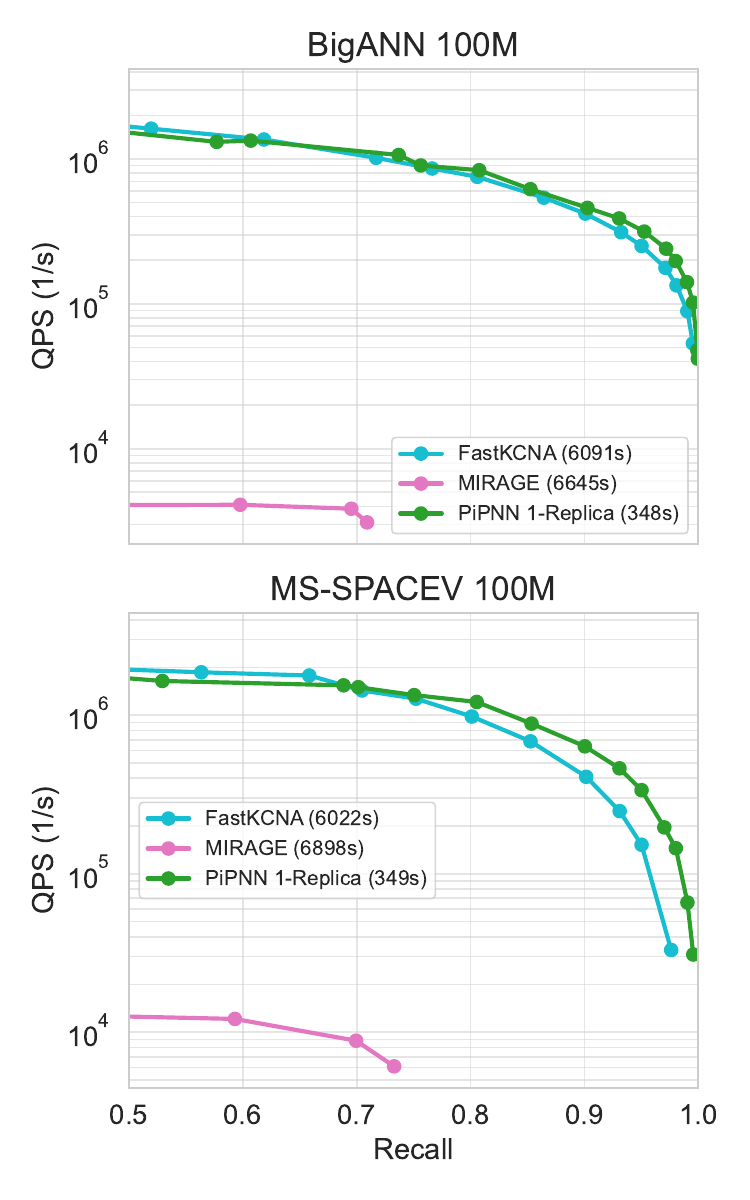}
    \caption{Results for the 100M sized ablation datasets MS-SPACEV-100M and BigANN-100M comparing \ourmethod{} against new recent methods for scaleable index build. }
    \label{fig:supp_newmethod_plot}
\end{figure}

We were unable to evaluate MIRAGE~\cite{voruganti2025mirage} and FastKCNA~\cite{yang2024revisiting} on billion scale datasets due to excessive memory usage. We had to modify MIRAGE to run on 100M datasets, as described in \Cref{sec:miragemods}, but did not need to require FastKCNA. We show results for our two 100M sized datasets used in our \ourmethod{} ablation in \Cref{fig:supp_newmethod_plot}. FastKCNA achieves slightly worse query performance than \ourmethod{} but with builds times at least $17.3\times$ longer. MIRAGE was unable to achieve competitive performance, and had build times at least $19.1\times$ longer then \ourmethod{}. 

\begin{figure}
    \centering
    \includegraphics[width=\linewidth]{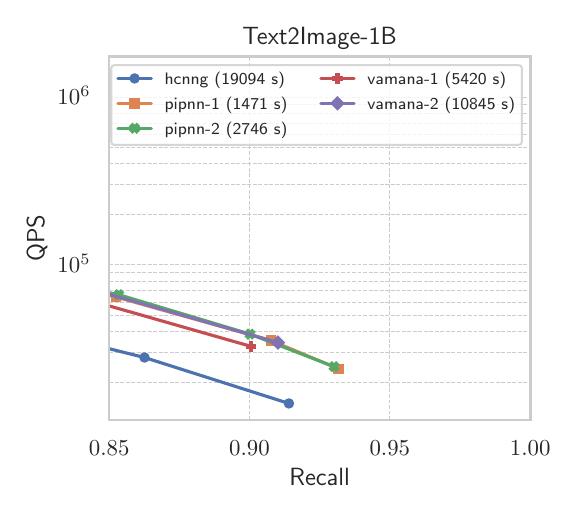}
    \caption{\ourmethod{} achieves the highest recall on the difficult Text2Image dataset.}
    \label{fig:t2i}
\end{figure}

\Cref{fig:t2i} shows results for the methods able to run the billion-scale Text2Image dataset~\cite{simhadri2022results} comprised of 200-dimensional float vectors outfitted with the MIPS dissimilarity measure. HNSW is not displayed due to being unable to achieve 0.8 recall. \ourmethod{} performs better than other methods with respect to maximum achieved recall. 

\begin{table}[t]
\centering
\small
\caption{\small{} Average hardware counters from \texttt{perf} per method and dataset. Values are shown in hundred billions $(\times 10^{11})$ for cycles, instructions, and LLC-load-misses. Pre-fetching is disabled for Vamana in this experiment. \ourmethod{} indexes datasets in fewer cycles and with fewer cache-misses.}
\label{tab:perf-averages}
\setlength{\tabcolsep}{6pt}
\resizebox{\columnwidth}{!}{%
\begin{tabular}{llrrrr}
\toprule
Method & Stat $(\times 10^{11})$ & bigann & openai & spacev & wiki \\
\midrule
\multirow{3}{*}{pipnn} & cycles & 935.71 & 85.76 & 927.19 & 1562.61 \\
 & instructions & 1170.70 & 90.99 & 1159.58 & 2066.17 \\
 & LLC-load-misses & 3.08 & 0.29 & 3.14 & 4.39 \\
\midrule
\multirow{3}{*}{vamana} & cycles & 5490.51 & 1109.12 & 3849.37 & 17180.30 \\
 & instructions & 5270.01 & 217.27 & 3445.63 & 3089.61 \\
 & LLC-load-misses & 11.83 & 3.05 & 9.84 & 63.99 \\
\midrule
\multirow{3}{*}{hnsw} & cycles & 4913.36 & 735.54 & 4381.94 & 8376.25 \\
 & instructions & 1098.43 & 92.08 & 1071.55 & 3849.41 \\
 & LLC-load-misses & 12.41 & 6.47 & 10.11 & 74.78 \\
\bottomrule
\end{tabular}
}
\end{table}

\begin{table}[!htbp]
\begin{tabular}{lcccc}
\toprule
Table Size & 64 & 128 & 192 & NoRes \\
Hash Bits &  &  &  &  \\
\midrule
8 & 3376.0 & 3545.0 & 3656.0 & 3445.0 \\
10 & 3433.0 & 3394.0 & 3614.0 & 3445.0 \\
12 & 3375.0 & 3559.0 & 3566.0 & 3445.0 \\
14 & 3531.0 & 3423.0 & 3421.0 & 3445.0 \\
16 & 3368.0 & 3653.0 & 3654.0 & 3445.0 \\
\bottomrule
\end{tabular}
\caption{Average cmps for Recall > 0.95 (SPACEV-100M Dataset)}
\label{tab:hashbit_data}
\end{table}

\subsection{Asymptotic Analysis of \ourmethod{}}

We now consider the expected running time to build a \ourmethod index. First, we point out that although in some adversarial input sets certain leader choices may lead to locally-unbalanced partitioning across subproblems, it is reasonable in practice to assume that randomized ball carving divides points across subproblems roughly evenly in expectation. Under this assumption, recursive ball carving with $\ell$ random leaders in each subproblem terminates within $O(\log_\ell n)$ levels with high probability, where $n$ is the number of input points. Within each level, up to $f n$ instances of points each undergo $d$-dimensional distance computations with the $\ell$ leaders in its subproblem (where $f$ is the amount of fanout performed at the prior level of recursion). It follows that partitioning takes $O(d f n \ell \log_\ell (n))$ time with high probability. 

We stop partitioning when every leaf contains at most $c$ points. We subsequently spend $O(|\mathcal{B}_i|^2d)$ time for each leaf $\mathcal{B}_i$ building a $|\mathcal{B}_i| \times |\mathcal{B}_i|$ distance matrix and adding edges between each point in $\mathcal{B}_i$ and the two points from which it has smallest distance. Thus, we can charge each point in $\mathcal{B}_i$ for up to $O(|\mathcal{B}_i|d) \leq O(cd)$ time. In total, there are $\leq f n$ instances of points across all leaves, so leaf-building in total takes $O(c d f n)$ time.

\HashPrune maintains a reservoir of up to $O(m)$ candidate neighbors for each point, where $m$ is the target maximum degree of the final graph. Recall from the leaf-building step that each of the $\leq f n$ instances of points adds a candidate edge to and from each of its two nearest in-leaf neighbors. Thus, at most $4fn$ candidate neighbors are passed into \HashPrune across all points. We hash each one via LSH against roughly 12 random vectors using 12 $O(d)$-time distance computations. We subsequently compare its distance from the target point with that of the furthest candidate in the reservoir (via brute force in $O(m)$ time) and the candidate already in its hash bucket if one exists (via binary search in $O(\log m)$ time). Thus, \HashPrune consumes $O(f n m)$ time.

Finally, having used \HashPrune to limit the number of candidates to $O(m)$ per point, \RobustPrune takes only $O(m^2)$ time for each point. Thus, the final prune phase takes $O(n m^2)$ time.

Combining all components of the algorithm produces a total expected time complexity of $O(n(d f \ell \log_\ell (n) + c d f + f m + m^2))$. 
\subsection{Proof of Determinism of \ourmethod{}}

The determinism of \ourmethod{} follows from two propositions. Firstly, the partitioning produced by Algorithm \ref{alg:Partition} is deterministic. Thus the candidates which are given to \HashPrune are always the same (though the order that they are merged into candidate lists by way of \HashPrune is non-deterministic). Secondly, \HashPrune itself is history independent, and so the order that points has no bearing on the final adjacency lists for each point. Thus \ourmethod{} is deterministic. We prove the each proposition in this section.

\begin{lemma}{Randomized Ball Carving is deterministic}
In each subproblem the leaders are sampled randomly (Line 3), thus once randomness is fixed the same leaders will always be sampled within a given subproblem. The next subproblems are given by computing an exact nearest neighbor computation on the leaders, and thus are completely determined by the leaders and subproblem points. Merging likewise can be done in such a way as to not depend on the scheduler. 
    
\end{lemma}

\begin{lemma}{\HashPrune is History Independent}\label{lem:hashprune-history-independence}

Let $p \subseteq \mathcal{X}$ be an arbitrary point, and consider $p$'s \HashPrune reservoir $R_p=[r_1,\dots,r_s]$ with hash family $H=\{h_1,\dots,h_b\}$. Suppose some fixed collection $C=\{c_1,\dots,c_i,\dots,c_q\}$ is to be inserted into $R_p$. We claim that $R_p$ will contain the $s$ elements of $C$ nearest to $p$ such that they do not collide under $H$. To see that this is true, consider that if we removed the size restriction of the reservoir entirely, then each position in $R_p$ would just contain the nearest point in each hash bucket (lines 3-5 of \HashPrune). Now, when we consider only the $s$-sized reservoir, the further member of $R_p$ is evicted in line 10 whenever the reservoir is full. Thus, the sequence of insertions is irrelevant to the final structure of $R_p$.

\end{lemma}

\subsection{Modifying MIRAGE}\label{sec:miragemods}

In our experiments, we compare against the MIRAGE nearest neighbor index. For transparency, we point out that, at the time of writing this, the native MIRAGE codebase does not currently support datasets much larger than the 10 million scale, because indices are done using signed 32-bit integers. Thus, a maximum degree of 128, for example, will lead to an integer overflow on the number of edges when the number of vertices exceeds 33.6 million. In order to let MIRAGE run on WikiCohere-35M, we modify the code to use unsigned 64-bit integers instead, which could impact the performance of the implementation. With that said, we see little noticeable effect on the build and query speeds of the graphs. Moreover, this aligns with all the other methods, which already represent point IDs in 64 bits.

\subsection{Additional Experiments}\label{sec:additional_experiments}

\begin{figure}
    \centering
    \includegraphics[width=\linewidth]{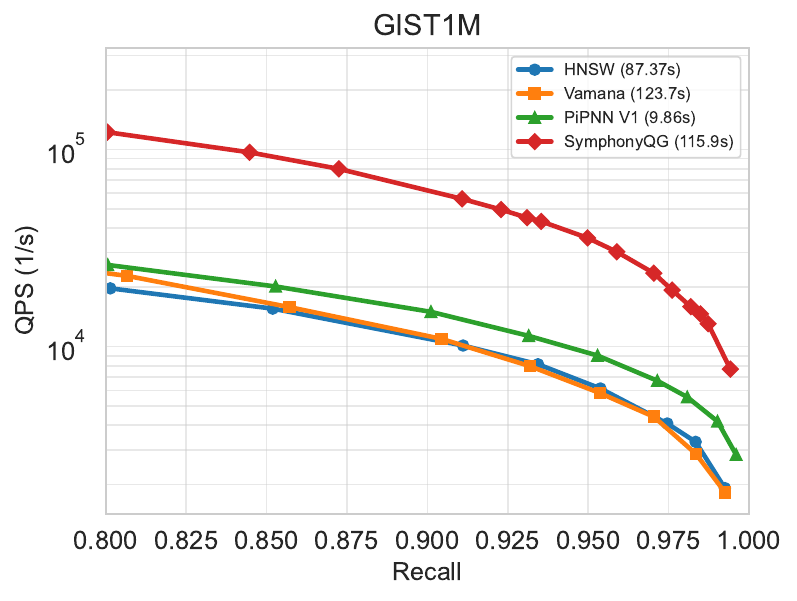}
    \caption{QPS-recall experiment on GIST-1M, a 960-dimension float32 dataset}
    \label{fig:gist_symphonyqg}
\end{figure}

\begin{figure}
    \centering
    \includegraphics[width=\linewidth]{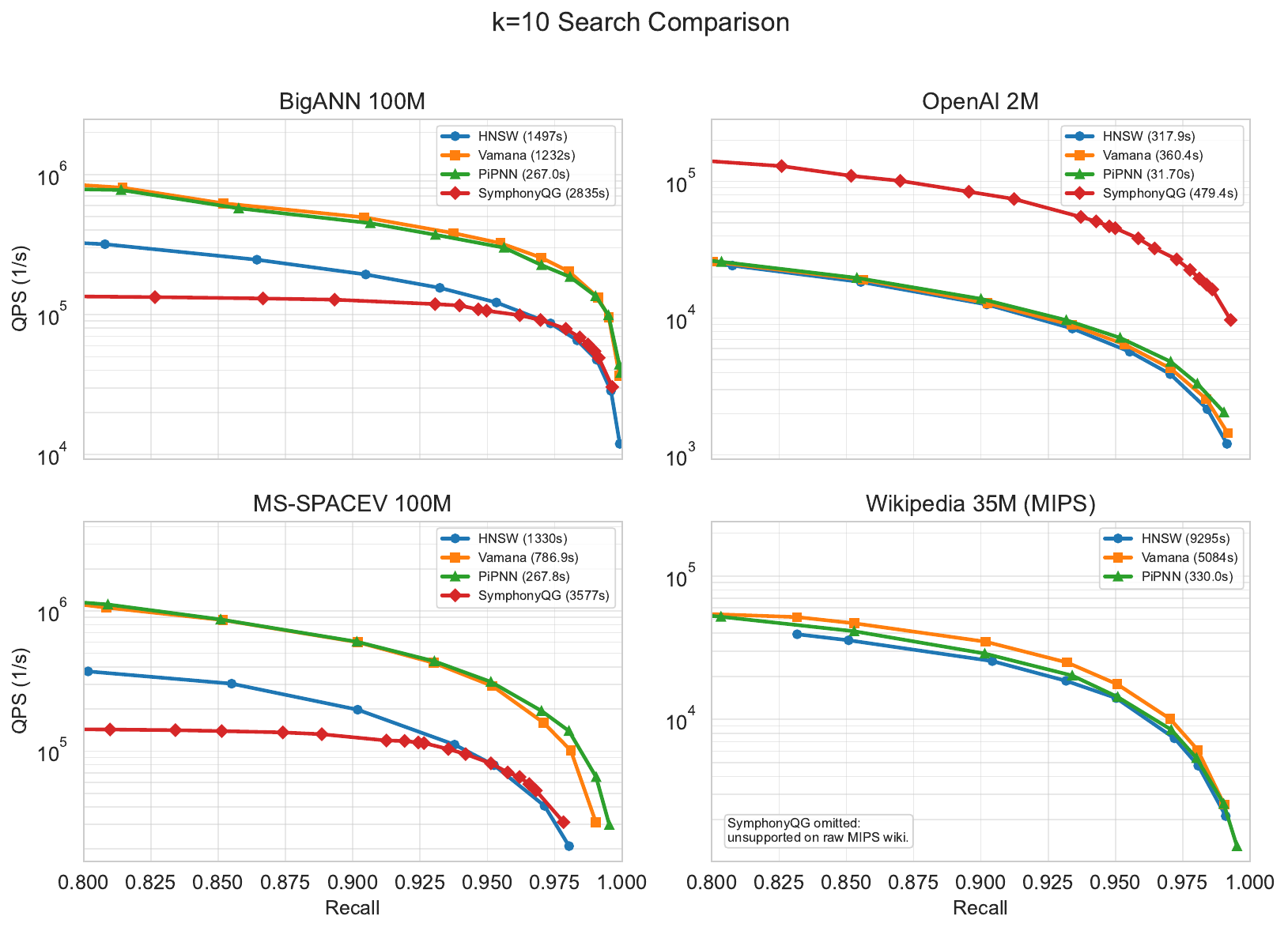}
    \caption{QPS-recall experiments on our ablation datasets including SymphonyQG as a baseline.}
    \label{fig:k10_symphonyqg}
\end{figure}

On request, we ran several additional experiments, the results of which we report here. In these, we include SymphonyQG~\cite{gou2025symphonyqg} as an additional baseline. SymphonyQG attaches RaBitQ codes to edges, which it uses to estimate distances during search. This allows SymphonyQG to avoid many distance computations on full-precision vectors, speeding up query times and potentially build times. \Cref{fig:k10_symphonyqg} shows QPS-recall curves on our ablation datasets, and \Cref{fig:gist_symphonyqg} shows additionally on the GIST dataset. 
We chose to restrict these experiments to our ablation datasets since SymphonyQG's memory requirements did not permit us to run it on our billion-scale datasets on our machine.
SymphonyQG's performance is shown on the three Euclidean datasets among these, as the implementation we used (from the RaBitQ library) does not appear to support inner product similarity.

On these datasets, we found that SymphonyQG takes longer to build than our other baselines but does manage to achieve much higher QPS on OpenAI-2M, attributable to the RaBitQ codes being used at query time. We point out that this is not an apples-to-apples comparison, as \ourmethod is optimized for build speed, whereas SymphonyQG is optimized for query speed. Moreover, all other methods we tested on have quantization disabled, due to varying levels of support for quantization. It does however, suggest an option for improving the query performance of other methods. Namely, the RaBitQ edge labels in SymphonyQG could readily be applied to any of the other graph indexes (\ourmethod included) as a post-processing step after construction, trading off some amount of build time in exchange for faster queries. Quantization can also easily be applied directly to base/query vectors to speed up distance comparisons. To this end, we direct the reader to our upcoming work evaluating quantization within \ourmethod as a method to increase both build and query speeds.

\myparagraph{Querying with larger $k$}

\begin{figure}
    \centering
    \includegraphics[width=\linewidth]{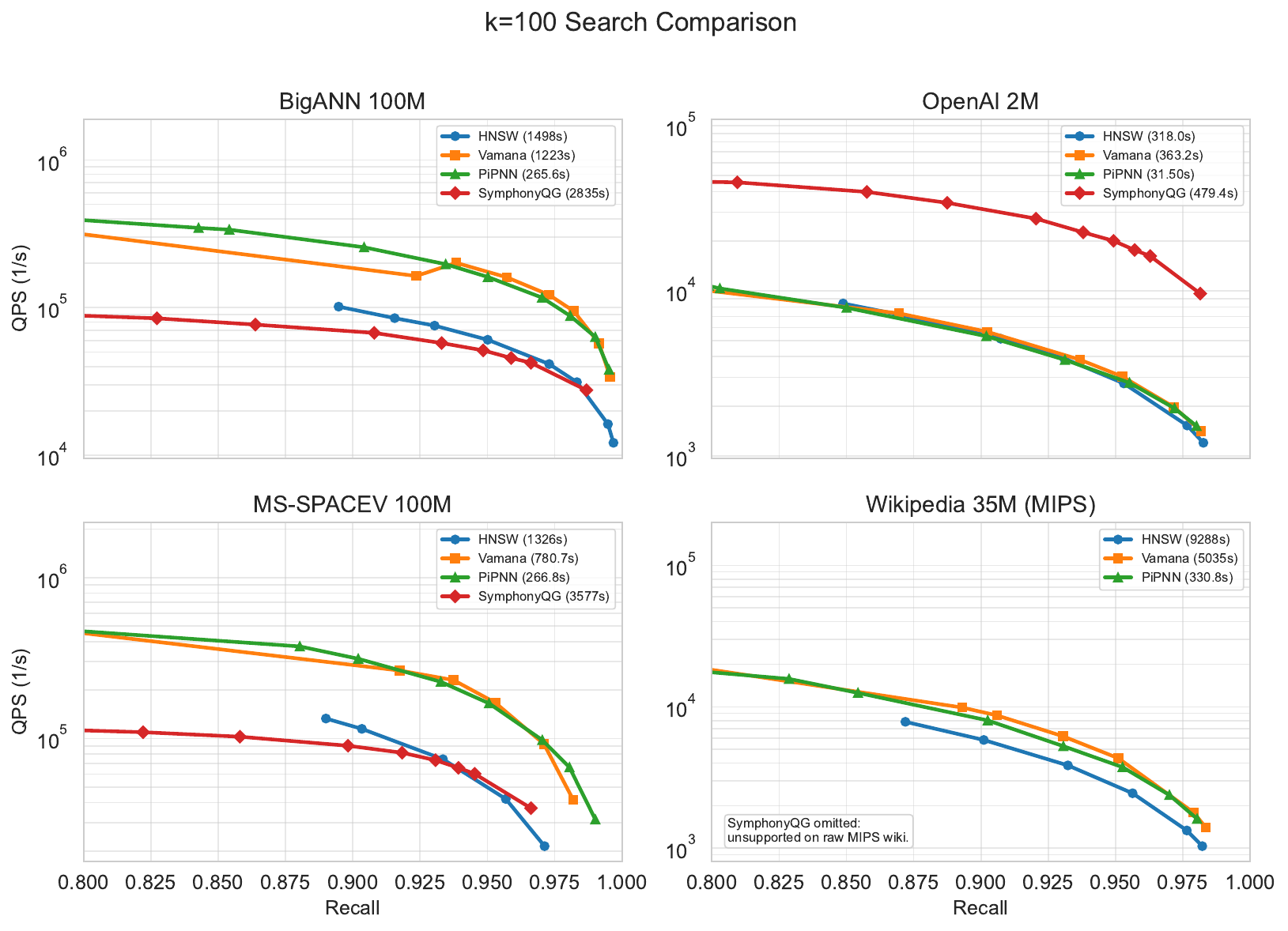}
    \caption{QPS-recall experiments on our ablation datasets with $k=100$.}
    \label{fig:k100_symphonyqg}
\end{figure}

In all of our recall experiments, we specifically measured $10@10$. Choosing $k=10$ is typical in ANNS literature and competitions. However, this may cause speculation as to whether \ourmethod is capable of maintaining search parity with state-of-the-art graph indexes at higher values of $k$. We repeated the QPS-recall experiments of \Cref{fig:k10_symphonyqg} but this time with $k=100$ (i.e. $100@100$ recall). As seen in \Cref{fig:k100_symphonyqg}, this has virtually no effect on the relative performance of the methods. \ourmethod maintains QPS-recall parity with the state-of-the-art. We did also run experiments with the more extreme $k=1000$, but all methods failed to achieve adequate recall.

\myparagraph{Memory footprint}

\begin{table*}[!htbp]
\begin{tabular}{lrrrrrr}
\toprule
Dataset & PiPNN & HNSW & Vamana & HCNNG & Mirage & SymQG \\
\midrule
BigANN 100M & 214.45 & 105.44 & 51.62 & 59.02 & 411.32 & 479.28 \\
MS-SPACEV 100M & 211.62 & 102.55 & 51.98 & 58.80 & 432.39 & 457.13 \\
OpenAI 2M & 18.01 & 28.66 & 17.17 & 24.79 & 33.98 & 47.37 \\
Wikipedia 35 M (MIPS) & 177.32 & 228.88 & 115.12 & 122.15 & 335.01 & NA \\
\bottomrule
\end{tabular}
\caption{Peak memory usage (GiB) during construction on ablation datasets with 128 threads.}
\label{tab:ablation-build-peak-memory}
\end{table*}

Here we report the total space used by each algorithm. We run each of the methods on our four ablation datasets and measure the total resident set size using the GNU time utility. The maximum memory footprint of the methods are reported in \Cref{tab:ablation-build-peak-memory} in gigabytes. \ourmethod uses notably more memory than Vamana and HNSW on our small integer datasets but comparable memory on our high-dimensional floating-point datasets. This is because \ourmethod uses significant overhead to store buckets during partitioning and leaf-building, but this overhead scales linearly with the number of points in the dataset and not its dimensionality or data type. Thus, the overhead is masked on higher-dimensional datasets, whereas it represents a larger fraction of the memory usage on small integer datasets.

\end{document}